\documentclass[12pt]{article}
\usepackage{graphicx}
\usepackage{amsfonts}
\usepackage{amssymb}
\usepackage{mathrsfs}
\usepackage{amsmath}
\usepackage{slashbox}
\begin{document}
\renewcommand{\thefootnote}{\fnsymbol{footnote}}
\begin{titlepage}

\vspace{10mm}
\begin{center}
{\Large\bf Maximally localized states and quantum corrections of black hole thermodynamics in the framework of a new generalized uncertainty principle}
\vspace{8mm}

{\large Yan-Gang Miao${}^{1,2,3,}$\footnote{E-mail address: miaoyg@nankai.edu.cn}, Ying-Jie Zhao${}^{1,}$\footnote{E-mail address: xiangyabaozhang@mail.nankai.edu.cn},  and Shao-Jun Zhang${}^{4,}$\footnote{E-mail address: sjzhang84@sjtu.edu.cn}

\vspace{6mm}
${}^{1}${\normalsize \em School of Physics, Nankai University, Tianjin 300071, China}

\vspace{3mm}
${}^{2}${\normalsize \em State Key Laboratory of Theoretical Physics, Institute of Theoretical Physics, \\Chinese Academy of Sciences, P.O. Box 2735, Beijing 100190, China}

\vspace{3mm}
${}^{3}${\normalsize \em Bethe Center for Theoretical Physics and Institute of Physics, University of Bonn, \\
Nussallee 12, D-53115 Bonn, Germany}

\vspace{3mm}
${}^{4}${\normalsize \em Center of Astronomy and Astrophysics, Shanghai Jiao Tong University, \\
Shanghai 200240, China}}

\end{center}

\vspace{4mm}
\centerline{{\bf{Abstract}}}
\vspace{4mm}
As a generalized uncertainty principle (GUP) leads to the effects of the minimal length of  the order of the Planck scale and UV/IR mixing, some significant physical concepts and quantities are modified or corrected correspondingly.
On the one hand, we derive the maximally localized states --- the physical states displaying the minimal length uncertainty  associated with a new  GUP proposed in our previous work. On the other hand, in the framework of this new GUP
we calculate quantum corrections to the thermodynamic quantities of the Schwardzschild black hole, such as the Hawking temperature, the entropy, and the heat capacity, and give a remnant mass of the black hole at the end of the evaporation process. Moreover, we compare our results with that obtained in the frameworks of several other GUPs. In particular, we observe a significant  difference between the situations  with and without the consideration of the UV/IR mixing effect in the quantum corrections to the evaporation rate and the decay time. That is, the decay time can greatly be prolonged 
in the former case, which implies that the quantum correction from the UV/IR mixing effect may give rise to a radical rather than a tiny influence to the Hawking radiation.

\vskip 10pt
\noindent
{\bf PACS Number(s)}: 02.40.Gh, 04.70.Dy, 03.65.Sq


\end{titlepage}
\newpage
\renewcommand{\thefootnote}{\arabic{footnote}}
\setcounter{footnote}{0}
\setcounter{page}{2}

\section{Introduction}
To unify general relativity and quantum mechanics is one of the most difficult tasks because the existing quantum gravity theories are ultraviolet divergent and thus non-renomaliziable. Various candidates of quantum gravity, including string theory~\cite{s1,s2,s3,s4}, loop quantum gravity~\cite{s5}, and quantum geometry~\cite{s6}, have pointed out that it is essential to introduce a fundamental length scale of the order of the Planck length and then the corresponding momentum provides a natural UV cutoff. Furthermore, Gedanken experiments of black holes~\cite{s7} tend to support the existence of a minimal length. One of the approaches to introduce a fundamental length scale is to modify the Heisenberg uncertainty principle (HUP) and then to obtain the so-called generalized uncertainty principle (GUP)~\cite{s8,s9,s10,s11,s12} whose commutation relations between position and momentum operators on a Hilbert space are no longer constants but depend in general on position and momentum operators. In the HUP framework, the restriction upon the position measurement precision does not exist. On the contrary, in the GUP framework that can be regarded as a phenomenological description of quantum gravity effects, a minimal position measurement precision is predicted with the order of the Planck length ${\ell_{\rm Pl}} = \sqrt {{{G\hbar }}/{{{c^3}}}}\sim 10^{-33}$ {\em cm} below which the spacetime cannot be probed effectively~\cite{s10,s11,s12,s13}. In other words, a finite resolution appears in the spacetime.

The idea of GUP has been utilized to modify fundamental physical concepts and to analyze the gravity effects on  fundamental physical quantities, such as maximally localized states~\cite{s8,Detournay} and energy spectra and wavefunctions of some interesting quantum systems~\cite{s14,s15,DasVagenas}, where the physical states displaying the minimal length uncertainty and the quantum corrections to energy spectra and wavefunctions have been calculated. 

On the other hand, the recent applications of GUP to the investigation of quantum black holes have attracted much attention and several achievements have been made~\cite{s16}. For instance, according to Hawking's black hole thermodynamics~\cite{Hawking}, a small black hole can radiate continuously and the black hole temperature can rise infinitely during the whole evaporation process until the black hole mass decreases to zero. However, in the framework of GUP the minimal length scale provides a natural restriction that the mass of a black hole cannot be less than the scale of the Planck mass at the end of the evaporation process, and the black hole remnant at the final stage of evaporation has zero entropy, zero heat capacity, and a finite temperature. Moreover, the entropy of a black hole does not strictly obey the area theorem but contains an additional logarithmic correction.

There exist some typical forms of GUP that give rise to modifications of basic concepts in quantum mechanics and to quantum gravity effects on black holes. Here we merely mention two of them that are intimately related to the present paper.
One is called the quadratic form~\cite{s8} (noted by $\mathrm{GUP_0}$ in the following figures, figure captions, and tables  for the sake of a concise presentation) in which the commutators of position and momentum operators contain an additional quadratic term of momentum operator. The other is called the exponential form~\cite{s17} (noted by $\mathrm{GUP_1}$ for the same purpose as $\mathrm{GUP_0}$) in which the commutators depend on an exponential function of the square of momentum operator. In fact, the former is just the first order approximation of the Taylor expansion of the latter in the Planck length. Based on the quadratic GUP, the maximally localized states are derived~\cite{s8} and then developed~\cite{Detournay} for a class of quite general GUPs. In the framework of the exponential GUP, the quantum corrections to the thermodynamic quantities of the Schwardzschild black hole are computed and some interesting results related to the black hole evaporation process are obtained, such as the faster evaporation and larger remnant mass than that deduced in the framework of the quadratic GUP.


In the present paper we revisit the maximally localized states and the quantum corrections to the thermodynamic quantities of the Schwardzschild black hole in the framework of our newly proposed GUP~\cite{s18}, the so-called improved exponential GUP (noted by $\mathrm{GUP_n}$ for the same purpose as $\mathrm{GUP_0}$ and $\mathrm{GUP_1}$). The motivation emerges directly from our recent interpretation that the origin of the cosmological constant problem  may arise from the GUP issue. Through choosing a suitable index $n$ introduced in our GUP and considering the UV/IR mixing effect, we can give the cosmological constant that coincides exactly with the experimental value provided by the most recent Planck 2013 results~\cite{Planck}. We are curious about how the maximally localized states are modified and how the thermodynamic quantities of the Schwardzschild black hole are corrected in the framework of our specific GUP. Following the scenario proposed in ref.~\cite{Detournay}, we obtain for our GUP the maximally localized states in terms of special functions. On the other hand, besides the expected outcomes that the corrected Hawking temperature, entropy, and heat capacity are distinct from that in the frameworks of other GUPs, our significant consequences lie on the two observations: One is that the evaporation rate is extremely small, in other words, the lifetime of black holes is remarkably prolonged, when the UV/IR mixing effect is particularly considered,  and the other observation is that the larger the index $n$ is, the less radiation the Schwardzschild black hole emits. 


The paper is arranged as follows. In the next section, we briefly review our improved exponential GUP with a particularly introduced positive integer $n$, give its minimal length and corresponding momentum, and then derive the maximally localized states. Based on our GUP, we work out in section 3 the corrected Hawking temperature, entropy, and heat capacity of the Schwarzschild black hole, and compare our results with that computed in the frameworks of the Hawking proposal, the quadratic GUP, and the exponential GUP.  We then turn to the Hawking evaporation process of the Schwarzschild black hole and calculate the quantum corrections to the evaporation rate and the decay time in section 4,  where we focus on the significant difference between the situations  without and with the consideration of the UV/IR mixing effect. 
Finally, we make a brief conclusion in section 5.

\section{The improved exponential GUP and its corresponding maximally localized states}  

\subsection{Representation of operators and the minimal length}
In ref.~\cite{s18} we propose our improved exponential GUP as follows,
\begin{eqnarray}
[ {\hat{X}, \hat{P}} ] = i\hbar \exp \left( \frac{{{\alpha ^{2n}}\ell_{\rm Pl}^{2n}}}{{{\hbar ^{2n}}}}{{\hat{P}^{2n}}} \right), \label{XandP}
\end{eqnarray}
where  $\alpha$ is a dimensionless parameter with the order of unity that describes the strength of gravitational effects,
and $n$ is a positive integer. Note that the parameter $\beta$ introduced in our original form, see ref.~\cite{s18}, has been set to be ${\alpha^{2}}\ell_{\rm Pl}^{2}/{\hbar^{2}}$ in order for us to make a direct comparison with the exponential GUP~\cite{s17} which is only our special case for $n=1$.
We point out that ${\alpha^{2}}\ell_{\rm Pl}^{2}/{\hbar^{2}}$ is very small  due to $(\ell_{\rm Pl}/ \hbar)^2 = (M_{\rm Pl}\,c)^{-2} \approx (10^{19}\, {\rm GeV})^{-2}$ when $\alpha$ is taken to be the order of unity in our discussion of micro black holes. Therefore, the deviation of our GUP from the HUP is kept small because the momentum of a particle is less than the Planck scale even if it is relatively large in some sense, which can be seen obviously from the Taylor expansion of our GUP. For phenomena at the other energy scales much less than the Planck one, such as those analyzed in refs.~\cite{s15,scar},
 $\alpha$ can have a large upper bound. As $\alpha$ being unity corresponds to the phenomena with momenta less than the Planck scale but much larger than that of those phenomena investigated, for instance, in refs.~\cite{s15,scar}, our setup of $\alpha$ has no conflict with the present experimental data.


In the momentum space the position and momentum operators can be represented as
\begin{eqnarray}
\hat{X} \psi (p) &=& i\hbar \exp \left( {\frac{{{\alpha ^{2n}}\ell_{\rm Pl}^{2n}}}{{{\hbar ^{2n}}}}{p^{2n}}} \right){\partial _p}\psi (p),
\label{Xp}\\
\hat{P} \psi (p) &=& p\,\psi (p), \label{Pp}
\end{eqnarray}
and the symmetric condition~\cite{s8}, 
\begin{eqnarray}
\left( {\left\langle \phi  \right|\hat{X}} \right)\left| \psi  \right\rangle  = \left\langle \phi  \right|\left( {\hat{X}\left| \psi  \right\rangle } \right), \qquad
\left( {\left\langle \phi  \right|\hat{P}} \right)\left| \psi  \right\rangle  = \left\langle \phi  \right|\left( {\hat{P}\left| \psi  \right\rangle } \right),
\end{eqnarray}
gives rise to the following scalar product of wavefunctions and the orthogonality and completeness of eigenstates,
\begin{eqnarray}
\left\langle \phi | \psi \right\rangle  &=& \int_{ - \infty }^{ + \infty } {dp \,\exp \left( { - \frac{{{\alpha ^{2n}}\ell_{\rm Pl}^{2n}}}{{{\hbar ^{2n}}}}{p^{2n}}} \right){\phi ^*}(p)\,\psi (p)}, \\
\left\langle p | p' \right\rangle  &=& \exp \left( {\frac{{{\alpha ^{2n}}\ell_{\rm Pl}^{2n}}}{{{\hbar ^{2n}}}}{p^{2n}}} \right)\delta \left( {p - p'} \right), \\
1 &=& \int_{ - \infty }^{ + \infty } { dp\, \exp \left( { - \frac{{{\alpha ^{2n}}\ell_{\rm Pl}^{2n}}}{{{\hbar ^{2n}}}}{p^{2n}}} \right) {| p \rangle } {\langle  p |}},
\end{eqnarray}
where $|p \rangle$ and $|p' \rangle$ mean momentum eigenstates and $\psi (p) \equiv \langle p | \psi \rangle$ stands for  a wavefunction in the momentum space.


From eq.~(\ref{XandP}) we get the uncertainty relation,
\begin{eqnarray}
\left(\Delta X\right) \left(\Delta P\right) \ge \frac{\hbar }{2}\left\langle \exp \left(\frac{{{\alpha ^{2n}}\ell_{\rm Pl}^{2n}}}{{{\hbar ^{2n}}}}{{\hat{P}^{2n}}}\right) \right\rangle.\label{uncertainty1}
\end{eqnarray}
In light of the properties $\langle \hat{P}^{2n} \rangle \ge \langle \hat{P}^{2} \rangle^n$ and $\langle {{\hat{P}^{2}}}\rangle={{\langle \hat P\rangle^2} + \left(\Delta P\right)^2}$, we reduce the uncertainty relation to be
\begin{eqnarray}
\left(\Delta X\right) \left(\Delta P\right) \ge \frac{\hbar }{2}\exp\left\{\frac{{{\alpha ^{2n}}\ell_{\rm Pl}^{2n}}}{{{\hbar ^{2n}}}}\left( {{\langle \hat P\rangle^2} + \left(\Delta P\right)^2} \right)^n\right\}.\label{uncertainty2}
\end{eqnarray}

For simplicity, we take $\langle \hat{P} \rangle  = 0$. By using the definition of the Lambert $W$ function~\cite{s20}, we write the saturate uncertainty relation as follows,
\begin{eqnarray}
W\left( u \right)\exp \left( {W\left( u \right)} \right) = u,\label{sur}
\end{eqnarray}
where we have set up $W\left( u \right) \equiv   - 2n{\left( {\frac{{\alpha \ell_{\rm Pl}}}{\hbar }} \right)^{2n}}\left(\Delta P\right)^{2n}$ and $u \equiv  - 2n{\left( {\frac{{\alpha  \ell_{\rm Pl}}}{{2\,\Delta X}}} \right)^{2n}}$. The Lambert function remains single-valued when it is restricted to be not less than $-1$ in the range $- \frac{1}{e} \le u \le  0$.  As a result, it is straightforward to give the minimal length from eq.~(\ref{sur}),
\begin{eqnarray}
\left( {\Delta X} \right)_0 = \frac{{\alpha \ell_{\rm Pl}}}{2}{\left( {2ne} \right)^{\frac{1}{{2n}}}},
\end{eqnarray}
and its corresponding momentum measurement precision,
\begin{eqnarray}
{\left( {\Delta P} \right)_{\rm Crit}}= {\left( {\frac{1}{{2n}}} \right)^{\frac{1}{{2n}}}}\frac{\hbar }{{\alpha \ell_{\rm Pl}}}, \label{Cri}
\end{eqnarray}
which can also be regarded as the critical value to distinguish the sub- and trans-Planckian modes~\cite{s19}.

We make two comments on the minimal length and the critical momentum. The first is that ${\left( {\Delta P} \right)_{\rm Crit}}$ is certainly in the order of the Planck momentum, ${\left( {\Delta P} \right)_{\rm Crit}} \sim P_{\rm Pl}=M_{\rm Pl}\,c$, when $\left(\Delta X\right)_0$ is in the order of the Planck length, $\left(\Delta X\right)_0 \sim \ell_{\rm Pl}$. The second comment that further demonstrates the minimal length and the critical momentum is that the minimal length never goes to a macroscopic order of magnitude even for a quite great $n$, like  $n \sim 10^{123}$, see ref.~\cite{s18}. That is,  the minimal length is always around the Planck length and the critical momentum is always around the Planck momentum for any $n$, which gives a good property for our improved exponential GUP.

At the end of this subsection we solve eq.~(\ref{sur}) and give the momentum measurement precision in terms of the position measurement precision for our use in section 3,
\begin{eqnarray}
\Delta P = \frac{\hbar }{{2\,\Delta X}}\exp \left\{- \frac{1}{{2n}}W\left( { - 2n{{\left( {\frac{{\alpha \ell_{\rm Pl}}}{{2\,\Delta X}}} \right)}^{2n}}} \right) \right\}. \label{pmp}
\end{eqnarray}


\subsection{Functional analysis of the position operator}
The eigenvalue equation for the position operator in the momentum space in the framework of $\mathrm{GUP_n}$ is given by
\begin{eqnarray}
i\hbar \exp \left( {\frac{{{\alpha ^{2n}}\ell _{{\rm{Pl}}}^{2n}}}{{{\hbar ^{2n}}}}}{p^{2n}} \right){\partial _p}{\psi _\lambda }(p) = \lambda \,{\psi _\lambda}(p),
\end{eqnarray}
and the wavefunctions, i.e. the position eigenfunctions can be obtained by solving the above equation,
\begin{eqnarray}
{\psi_\lambda }(p) = \begin{cases} \sqrt {\frac{{\alpha {\ell _{{\rm{Pl}}}}}}{{2\hbar \Gamma \left( {\frac{{2n + 1}}{{2n}}} \right)}}} \exp \left\{ +{ i\lambda }\left[ {\frac{1}{{\alpha {\ell _{{\rm{Pl}}}}}}\Gamma \left( {\frac{{2n + 1}}{{2n}}} \right) + \frac{p}{{2n}{\hbar}}{E_{\frac{{2n - 1}}{{2n}}}}\left( {\frac{{{\alpha ^{2n}}\ell _{{\rm{Pl}}}^{2n}}}{{{\hbar ^{2n}}}}} {p^{2n}}\right)} \right] \right\}, &\mbox{{\rm if}\; $p < 0$},\\
\sqrt {\frac{{\alpha {\ell _{{\rm{Pl}}}}}}{{2\hbar \Gamma \left( {\frac{{2n + 1}}{{2n}}} \right)}}} \exp \left\{ -{ i\lambda }\left[ {\frac{1}{{\alpha {\ell _{{\rm{Pl}}}}}}\Gamma \left( {\frac{{2n + 1}}{{2n}}} \right) - \frac{p}{{2n}{\hbar}}{E_{\frac{{2n - 1}}{{2n}}}}\left( {\frac{{{\alpha ^{2n}}\ell _{{\rm{Pl}}}^{2n}}}{{{\hbar ^{2n}}}}} {p^{2n}}\right)} \right] \right\}, &\mbox{{\rm if}\; $p \ge 0$},
\end{cases}
\end{eqnarray}
where $\Gamma(x)$ is the gamma function defined as $\Gamma(x) \equiv \int_0^\infty  {{t^{x - 1}}{e^{ - t}}dt}$ and ${E_n}(x)$ is the generalized exponential integral function defined as ${E_n}(x) \equiv \int_1^\infty  {t^{ - n}} {e^{ - xt}}dt$. 
Note that this piecewise-defined function is continuous at the point $p=0$.

The scalar product of wavefunctions can be calculated,
\begin{eqnarray}
\langle{\psi _{\lambda'} }|{\psi _\lambda }\rangle &=& \int_{ - \infty }^{ + \infty } {dp}\, \exp \left( {-\frac{{ {\alpha ^{2n}}\ell _{{\rm{Pl}}}^{2n}}}{{{\hbar ^{2n}}}}}{p^{2n}} \right)\psi _{\lambda'} ^*(p)\,{\psi _\lambda }(p)  \nonumber \\
&=& \frac{1}{\left[{\frac{{\Gamma \left( {\frac{{2n + 1}}{{2n}}} \right)}}{{\alpha {\ell _{{\rm{Pl}}}}}}\left( {\lambda  - \lambda' } \right)}\right]}\sin \left[ {\frac{{\Gamma \left( {\frac{{2n + 1}}{{2n}}} \right)}}{{\alpha {\ell _{{\rm{Pl}}}}}}\left( {\lambda  - \lambda' } \right)} \right].
\end{eqnarray}
Note that, according to KMM's result~\cite{s8}, because of the existence of the minimal length there are no exact eigenvalues for the position operator and the formal eigenfunctions ${\psi_\lambda }(p) $ attained by solving the eigenvalue equation are in fact unphysical. For this reason we have to recover information on position by using the maximally localized states which will be analyzed below.

\subsection{Maximally localized states}
In order to recover information on position the maximally localized states are introduced and used to calculate the average values of the position operator instead of the ordinary position eigenvalues. In ref.~\cite{s8} the maximally localized states are constructed from the squeezed states satisfying
\begin{eqnarray}
\left(\Delta X\right) \left(\Delta P\right) = \frac{1}{2}\left| {\left\langle {[ {\hat{X},\hat{P}} ]} \right\rangle } \right|.
\end{eqnarray}
However, it is pointed out~\cite{Detournay} that only for a very special GUP, like the quadratic form $\mathrm{GUP_0}$,  can the maximally localized states be obtained in terms of squeezed states. In general, a constrained variational principle should be applied in order to find out maximally localized states. The states are solutions of the following Euler-Lagrange equation in the momentum space~\cite{Detournay},
\begin{eqnarray}
\left\{ { - {{\left[ {f(p){\partial _p}} \right]}^2} - {\xi ^2} + 2a\left[ {if(p){\partial _p} - \xi } \right] + 2b\left[ {v(p) - \gamma } \right] - {\mu ^2}} \right\}\Psi (p) = 0,\label{ELE}
\end{eqnarray}
where $a$ and $b$ are Lagrange multipliers, the function $f(p)$ depends on the commutator $[ {\hat{X},\hat{P}} ] = i f( \hat{P})$, and the other parameters emerge from the following relations,
\begin{eqnarray}
\left( {\Delta X} \right)_{\rm min}^2 = \min \frac{{\langle \Psi |{{\hat X}^2} - {\xi ^2}| \Psi\rangle}}{{\langle \Psi|\Psi \rangle}} \equiv {\mu ^2}, \qquad \xi \equiv \frac{{\langle \Psi |{\hat X}| \Psi\rangle}}{{\langle \Psi|\Psi \rangle}}, \qquad \gamma \equiv \frac{{\langle \Psi |v({\hat P}) | \Psi\rangle}}{{\langle \Psi|\Psi \rangle}}.\label{threepara}
\end{eqnarray}
Note that $v({\hat P})$ is such an operator that its representing function in the momentum space $v(p)$ diverges and is not integrable, but cannot diverge faster than $| p |^{3\nu}$ with $\nu > 0$ when $|p|$ goes to infinity. However, it is not necessary to determine the concrete form of  $v(p)$ as the maximally localized states appear under the condition $b=0$. For the details of relevant analysis, see ref.~\cite{Detournay}.

Furthermore, according to the proposal in ref.~\cite{Detournay}, when $|p| \rightarrow \infty$,  if $z(p)$ defined as
\begin{eqnarray}
z(p) \equiv \int_0^p {\frac{{dp'}}{{f(p')}}}
\end{eqnarray}
has finite limits,
\begin{eqnarray}
z({+\infty}) \equiv {\alpha _ + } > 0, \qquad z({-\infty}) \equiv {\alpha _ - } < 0,
\end{eqnarray}
one can solve the Euler-Lagrange equation (eq.~(\ref{ELE})) for $b=0$ and give the maximally localized states as follows,
\begin{eqnarray}
\Psi _\xi (p) \equiv \langle p | \Psi _\xi \rangle = C \exp [{ - i\xi z(p)}] \sin \{ {\mu [ {z(p) - {\alpha _ - }}]} \},\label{maxlocsta}
\end{eqnarray}
where
\begin{eqnarray}
| C | = \sqrt {\frac{2}{\hbar\left({\alpha _ + } - {\alpha _ - }\right)}} , \qquad \mu  = \frac{{k\pi }}{{{\alpha _ + } - {\alpha _ - }}}, \qquad k \in \mathbb{N}.
\end{eqnarray}
Correspondingly, the minimal spread in position for $k=1$ equals
\begin{eqnarray}
{\left( {\Delta X} \right)_{\min }}{\bigg|}_{b=0} = \frac{\pi }{{{\alpha _ + } - {\alpha _ - }}}.\label{minilength}
\end{eqnarray}

Now we turn to our case in which for the improved exponential GUP, $f(p)$ has the form,
\begin{eqnarray}
f( p ) = \hbar\exp \left( \frac{{\alpha ^{2n}}\ell_{\rm Pl}^{2n}}{{\hbar ^{2n}}}{p^{2n}} \right).
\end{eqnarray}
We thus calculate
\begin{eqnarray}
z(p) \equiv  \int_0^p {\frac{{dp'}}{{f(p')}}} = \begin{cases}  - \frac{1}{{\alpha \ell_{\rm Pl}}}\Gamma \left( {\frac{{2n + 1}}{{2n}}} \right) - \frac{p}{{2n\hbar }}{E_{\frac{{2n - 1}}{{2n}}}}\left( \frac{{\alpha ^{2n}}\ell_{\rm Pl}^{2n}}{{\hbar ^{2n}}}{p^{2n}} \right), &\mbox{{\rm if}\; $ p < 0$},\\
+ \frac{1}{{\alpha \ell_{\rm Pl}}}\Gamma \left( {\frac{{2n + 1}}{{2n}}} \right) - \frac{p}{{2n\hbar }}{E_{\frac{{2n - 1}}{{2n}}}}\left( \frac{{\alpha ^{2n}}\ell_{\rm Pl}^{2n}}{{\hbar ^{2n}}}{p^{2n}} \right), &\mbox{{\rm if}\; $ p \ge 0$},\label{ourzp}
\end{cases}
\end{eqnarray}
and the finite parameters $\alpha _ + $ and $\alpha _ - $, respectively,
\begin{eqnarray}
{\alpha _ + } = \frac{1}{{\alpha \ell_{\rm Pl}}}\Gamma \left( {\frac{{2n + 1}}{{2n}}} \right), \qquad {\alpha _ - } = - \frac{1}{{\alpha \ell_{\rm Pl}}}\Gamma \left( {\frac{{2n + 1}}{{2n}}} \right).\label{alphapm}
\end{eqnarray}
As a result, we deduce the maximally localized states in the momentum space which can be written as a  piecewise-defined function. For $p < 0$, it can be expressed as
\begin{eqnarray}
\Psi _\xi (p) &=& -\sqrt {\frac{{\alpha \ell_{\rm Pl}}}{{ \hbar\Gamma \left( {\frac{{2n + 1}}{{2n}}} \right)}}}
\exp \left[ { \frac{{i\xi \Gamma \left( {\frac{{2n + 1}}{{2n}}} \right)}}{{\alpha \ell_{\rm Pl}}}}
+ {\frac{{i\xi}}{{2n\hbar }}\,{p\,E_{\frac{{2n - 1}}{{2n}}}}\left( \frac{{\alpha ^{2n}}\ell_{\rm Pl}^{2n}}{{\hbar ^{2n}}} {p^{2n}}\right)} \right]  \nonumber \\
& & \times \sin \left[ {\frac{{\pi \alpha \ell_{\rm Pl}}}{{4n\hbar \Gamma \left( {\frac{{2n + 1}}{{2n}}} \right)}}\,{p\,E_{\frac{{2n - 1}}{{2n}}}}\left( \frac{{\alpha ^{2n}}\ell_{\rm Pl}^{2n}}{{\hbar ^{2n}}} {p^{2n}}\right)} \right],
\end{eqnarray}
and for $ p \ge 0$ as
\begin{eqnarray}
\Psi _\xi (p) &=& \sqrt {\frac{{\alpha \ell_{\rm Pl}}}{{ \hbar\Gamma \left( {\frac{{2n + 1}}{{2n}}} \right)}}}
\exp \left[ { - \frac{{i\xi \Gamma \left( {\frac{{2n + 1}}{{2n}}} \right)}}{{\alpha \ell_{\rm Pl}}}}
+ {\frac{{i\xi}}{{2n\hbar }}\,{p\,E_{\frac{{2n - 1}}{{2n}}}}\left( \frac{{\alpha ^{2n}}\ell_{\rm Pl}^{2n}}{{\hbar ^{2n}}} {p^{2n}}\right)} \right]  \nonumber \\
& & \times \sin \left[ {\frac{{\pi \alpha \ell_{\rm Pl}}}{{4n\hbar \Gamma \left( {\frac{{2n + 1}}{{2n}}} \right)}}\,{p\,E_{\frac{{2n - 1}}{{2n}}}}\left( \frac{{\alpha ^{2n}}\ell_{\rm Pl}^{2n}}{{\hbar ^{2n}}} {p^{2n}}\right)} \right].
\end{eqnarray}
The minimal length that corresponds to the maximally localized states then reads as
\begin{eqnarray}
{\left( {\Delta X} \right)_{\min }}{\bigg|}_{b=0} = \frac{{\pi \alpha \ell_{\rm Pl}}}{{2\Gamma \left( {\frac{{2n + 1}}{{2n}}} \right)}}.
\end{eqnarray}

In the remaining contexts of this subsection, we give some interesting properties of the maximally localized states.

First of all, we point out that any two maximally localized states with different positions $\xi$'s (see eq.~(\ref{threepara})) are no longer mutually orthogonal because of the fuzziness of position space,
\begin{eqnarray}
\langle\Psi _{\xi'} | \Psi _{\xi}\rangle 
= \frac{{\pi ^2}{\alpha ^3}{\ell_{\rm Pl}^3}}{ \Gamma \left( {\frac{{2n + 1}}{{2n}}} \right)}\,
\frac{\sin \left[{\frac{{\Gamma \left( {\frac{{2n + 1}}{{2n}}} \right)}}{{\alpha \ell_{\rm Pl}}}}\left( {\xi  - \xi '} \right)\right]} {{\pi ^2}{\alpha ^2}\ell_{\rm Pl}^2\left( {\xi  - \xi '} \right) - \left[\Gamma\left( {\frac{{2n + 1}}{{2n}}} \right)\right]^2{{\left( {\xi  - \xi '} \right)}^3}}.\label{nonorth}
\end{eqnarray}

Next, we project an arbitrary state $| \psi \rangle$ onto one maximally localized state and calculate the probability amplitude for the particle being maximally localized around the position $\xi$. To this end, we write the transformation of a wavefunction from the momentum space to the quasi-position space,
\begin{eqnarray}
\psi(\xi) &\equiv& \langle \Psi _\xi  | \psi \rangle    \nonumber \\
&=& \left\{-\sqrt {\frac{{\alpha \ell_{\rm Pl}}}{{ \hbar\Gamma \left( {\frac{{2n + 1}}{{2n}}} \right)}}} \exp \left[ {- \frac{{i\xi \Gamma \left( {\frac{{2n + 1}}{{2n}}} \right)}}{{\alpha \ell_{\rm Pl}}}} \right]\int_{ - \infty }^{0} +\sqrt {\frac{{\alpha \ell_{\rm Pl}}}{{ \hbar\Gamma \left( {\frac{{2n + 1}}{{2n}}} \right)}}} \exp \left[ {+ \frac{{i\xi \Gamma \left( {\frac{{2n + 1}}{{2n}}} \right)}}{{\alpha \ell_{\rm Pl}}}} \right]\int_{0}^{ + \infty }\right\}\nonumber \\
& & \times \, dp \,\Bigg\{\exp \left( -\frac{{\alpha ^{2n}}\ell_{\rm Pl}^{2n}}{{\hbar ^{2n}}}{p^{2n}} \right)\exp \left[{-\frac{{i\xi}}{{2n\hbar }}\,{p\,E_{\frac{{2n - 1}}{{2n}}}}\left( \frac{{\alpha ^{2n}}\ell_{\rm Pl}^{2n}}{{\hbar ^{2n}}}{p^{2n}} \right)} \right] \nonumber \\
& & \times \sin \left[ {\frac{{\pi \alpha \ell_{\rm Pl}}}{{4n\hbar \Gamma \left( {\frac{{2n + 1}}{{2n}}} \right)}}\,{p\,E_{\frac{{2n - 1}}{{2n}}}}\left( \frac{{\alpha ^{2n}}\ell_{\rm Pl}^{2n}}{{\hbar ^{2n}}} {p^{2n}}\right)} \right] \psi(p)\Bigg\}.\label{trans}
\end{eqnarray}
For instance, the quasi-position wavefunction\footnote{It can be obtained by simply substituting the momentum eigenfunction ${\psi _{\tilde{p}}}\left( p \right) = \delta \left( {p - {\tilde p}} \right)$ into eq.~(\ref{trans}).} of the momentum eigenfunction ${\psi _{\tilde{p}}}\left( p \right)$ with the eigenvalue ${\tilde p}$ is always a plane wave but has a specific wavelength,
\begin{eqnarray}
\lambda_\xi  = \frac{{2\pi }}{{ {\frac{1}{{ \alpha {\ell _{{\rm{Pl}}}}}}\Gamma \left( {\frac{{2n + 1}}{{2n}}} \right) - \frac{|\tilde{p}|}{{2n\hbar }}{E_{\frac{{2n - 1}}{{2n}}}}\left( {\frac{{{\alpha ^{2n}}\ell _{{\rm{Pl}}}^{2n}}}{{{\hbar ^{2n}}}}}{{\tilde{p}}^{2n}} \right)}}} > \frac{{2\pi \alpha {\ell _{{\rm{Pl}}}}}}{{\Gamma \left( {\frac{{2n + 1}}{{2n}}} \right)}},
\end{eqnarray}
which reveals that the physical states with wavelengths shorter than $\frac{{2\pi \alpha {\ell _{{\rm{Pl}}}}}}{{\Gamma \left( {\frac{{2n + 1}}{{2n}}} \right)}}$ are naturally discarded in the process of the generalized Fourier decomposition of the quasi-position wavefunction of physical states. However, we mention that in the ordinary quantum mechanics the position wavefunction describes a physical state and thus no physical states with short wavelengths are discarded in the Fourier decomposition. By using eq.~(\ref{trans}), we can calculate   the probability amplitude for the particle being maximally localized around the position $\xi$, which can be read out from the scalar product of quasi-position wavefunctions discussed below.

At last, we give the scalar product of quasi-position wavefunctions by first deriving
the inverse transformation of eq.~(\ref{trans}),
\begin{eqnarray}
\psi (p) &=&  \frac{1}{{2\pi }}\sqrt {\frac{{\Gamma \left( {\frac{{2n + 1}}{{2n}}} \right)}}{{\alpha \ell_{\rm Pl}\hbar}}}\,
\csc \left[ {\frac{{\pi \alpha \ell_{\rm Pl}}}{{4n\hbar\Gamma \left( {\frac{{2n + 1}}{{2n}}} \right)}}\,{p\,E_{\frac{{2n - 1}}{{2n}}}}\left( \frac{{\alpha ^{2n}}\ell_{\rm Pl}^{2n}}{{\hbar ^{2n}}}{p^{2n}} \right)} \right]      \nonumber \\
& &\times \,\Bigg\{-  \int_{ - \infty }^{0} d\xi \,\exp \left[ {\frac{{i\xi \Gamma \left( {\frac{{2n + 1}}{{2n}}} \right)}}{{\alpha \ell_{\rm Pl}}}}+{\frac{{i\xi}}{{2n\hbar}}\,{p\,E_{\frac{{2n - 1}}{{2n}}}}\left( \frac{{\alpha ^{2n}}\ell_{\rm Pl}^{2n}}{{\hbar ^{2n}}} {p^{2n}}\right)} \right]\psi(\xi) \nonumber \\
& &\hspace{8mm}+  \int_{0}^{ + \infty } d\xi \,\exp \left[ {-\frac{{i\xi \Gamma \left( {\frac{{2n + 1}}{{2n}}} \right)}}{{\alpha \ell_{\rm Pl}}}}+{\frac{{i\xi}}{{2n\hbar}}\,{p\,E_{\frac{{2n - 1}}{{2n}}}}\left( \frac{{\alpha ^{2n}}\ell_{\rm Pl}^{2n}}{{\hbar ^{2n}}} {p^{2n}}\right)}\right] \psi(\xi)\Bigg\},\label{invertrans}
\end{eqnarray}
and then computing the following integration,
\begin{eqnarray}
{\langle \phi  | \psi \rangle}
&=& \int_{ - \infty }^{ + \infty } {dp} \,\exp \left(- {\frac{{{\alpha ^{2n}}\ell _{{\rm{Pl}}}^{2n}}}{{{\hbar ^{2n}}}}} {p^{2n}}\right)  {\phi ^*}(p)\,\psi(p) \nonumber \\
 &=& \frac{{\Gamma \left( {\frac{{2n + 1}}{{2n}}} \right)}}{{4{\pi ^2}\alpha \ell_{\rm Pl}}\hbar} \int_{ - \infty }^{ + \infty } {dp} \,\Bigg\{\exp \left( {-\frac{{{\alpha ^{2n}}\ell _{{\rm{Pl}}}^{2n}}}{{{\hbar ^{2n}}}}}{p^{2n}} \right) {\csc ^2}\left[ {\frac{{\pi \alpha \ell_{\rm Pl}}}{{4n\hbar\Gamma \left( {\frac{{2n + 1}}{{2n}}} \right)}}\,{p\,E_{\frac{{2n - 1}}{{2n}}}}\left( \frac{{\alpha ^{2n}}\ell_{\rm Pl}^{2n}}{{\hbar ^{2n}}}{p^{2n}} \right)} \right]      \nonumber \\
& & \times \,\Bigg[-  \int_{ - \infty }^{0} d\xi^{\prime} \,\exp \left[ {-\frac{{i\xi^{\prime} \Gamma \left( {\frac{{2n + 1}}{{2n}}} \right)}}{{\alpha \ell_{\rm Pl}}}}-{\frac{{i\xi^{\prime}}}{{2n\hbar}}\,{p\,E_{\frac{{2n - 1}}{{2n}}}}\left( \frac{{\alpha ^{2n}}\ell_{\rm Pl}^{2n}}{{\hbar ^{2n}}} {p^{2n}}\right)} \right]\phi^*(\xi^{\prime}) \nonumber \\
& &\hspace{8mm}+  \int_{0}^{ + \infty } d\xi^{\prime} \,\exp \left[ {\frac{{i\xi^{\prime} \Gamma \left( {\frac{{2n + 1}}{{2n}}} \right)}}{{\alpha \ell_{\rm Pl}}}}-{\frac{{i\xi^{\prime}}}{{2n\hbar}}\,{p\,E_{\frac{{2n - 1}}{{2n}}}}\left( \frac{{\alpha ^{2n}}\ell_{\rm Pl}^{2n}}{{\hbar ^{2n}}} {p^{2n}}\right)}\right] \phi^*(\xi^{\prime})\Bigg]\nonumber \\
& &\times \,\Bigg[-  \int_{ - \infty }^{0} d\xi \,\exp \left[ {\frac{{i\xi \Gamma \left( {\frac{{2n + 1}}{{2n}}} \right)}}{{\alpha \ell_{\rm Pl}}}}+{\frac{{i\xi}}{{2n\hbar}}\,{p\,E_{\frac{{2n - 1}}{{2n}}}}\left( \frac{{\alpha ^{2n}}\ell_{\rm Pl}^{2n}}{{\hbar ^{2n}}} {p^{2n}}\right)} \right]\psi(\xi) \nonumber \\
& &\hspace{8mm}+  \int_{0}^{ + \infty } d\xi \,\exp \left[ {-\frac{{i\xi \Gamma \left( {\frac{{2n + 1}}{{2n}}} \right)}}{{\alpha \ell_{\rm Pl}}}}+{\frac{{i\xi}}{{2n\hbar}}\,{p\,E_{\frac{{2n - 1}}{{2n}}}}\left( \frac{{\alpha ^{2n}}\ell_{\rm Pl}^{2n}}{{\hbar ^{2n}}} {p^{2n}}\right)}\right] \psi(\xi)\Bigg]\Bigg\}.\label{quasiposiwf}
\end{eqnarray}

\section{Black hole thermodynamics}

In this section we calculate the quantum corrections to the Hawking temperature, the entropy, and the heat capacity of the Schwarzschild black hole in the framework of our GUP, and compare our results with that  obtained previously in the frameworks of the Hawking proposal, the quadratic GUP, and the exponential GUP.

In the following contexts of the present paper we adopt the units $\hbar = c = k_B =1$. As a result, the Planck length, the Planck mass, the Planck temperature, and the gravitational constant satisfy the relations: $\ell_{\rm Pl} = M_{\rm Pl}^{-1} =  T_{\rm Pl}^{-1} = \sqrt G $.

\subsection{Temperature}
The metric of a four-dimensional Schwarzschild black hole can be written as
\begin{eqnarray}
d{s^2} = - \left( {1 - \frac{{2MG}}{r}} \right)d{t^2} + {\left( {1 - \frac{{2MG}}{r}} \right)^{ - 1}}d{r^2} + {r^2}d{\Omega ^2},
\end{eqnarray}
where $M$ is the black hole mass.
The Schwarzschild horizon radius is defined as $r_{\rm h} \equiv 2MG$. According to the near-horizon geometry, the position measurement precision is of the order of the horizon radius,\footnote{We assume $\Delta X \simeq r_h$ as done in other works, see, for instance, ref.~\cite{s17}. This is a physical estimation and we think it is reasonable. For a static observer outside the horizon, one cannot fix the position of a particle around a black hole because of the horizon, so the coordinate-uncertainty of the particle can be estimated to be the radius of the horizon. This estimation (also including others) is from physical intuition and does not depend on the explicit form of a generalized uncertainty principle. Its validity can be confirmed from its successful inducing of the standard Hawking temperature of the Schwarzschild black hole.} $\Delta X \simeq r_{\rm h}$. Therefore, we can deduce that the minimal length corresponds to the minimal mass of the Schwarzschild black hole in such a way:  $\left( {\Delta X} \right)_0 \approx 2M_0G$, which gives the minimal mass, sometimes called the black hole remnant (BHR), as follows:
\begin{eqnarray}
{M_0} \approx \frac{{\alpha {M_{\rm Pl}}}}{4}{\left( {2ne} \right)^{\frac{1}{{2n}}}}. \label{Length}
\end{eqnarray}
Note that as $\alpha$ is in the order of unity the black hole remnant is of the order of the Planck mass for any $n$. 

Following the method~\cite{Scardigli} which connects directly the uncertainty relation with the black hole mass-temperature relation
and using eq.~(\ref{pmp}), we obtain the corrected temperature which is expressed in terms of the ratio of the minimal mass and the mass of the black hole,
\begin{eqnarray}
{T} \approx \frac{{\Delta P}}{{2\pi }} = \frac{1}{{8\pi MG}}\exp \left\{ { - \frac{1}{{2n}}W\left( { - \frac{1}{e}{{\left( {\frac{{{M_0}}}{M}} \right)}^{2n}}} \right)} \right\}. \label{TH}
\end{eqnarray}
In order to compare the above result with others, we expand it in $\frac{1}{e} \left( {{M_0}/M}\right)^{2n}$,
\begin{eqnarray}
{T} = \frac{1}{{8\pi MG}}\left[ {1 + \frac{1}{{2ne}}{{\left( {\frac{{{M_0}}}{M}} \right)}^{2n}} + \frac{{4n + 1}}{{8{n^2}{e^2}}}{{\left( {\frac{{{M_0}}}{M}} \right)}^{4n}} + \frac{{{{\left( {6n + 1} \right)}^2}}}{{48{n^3}{e^3}}}{{\left( {\frac{{{M_0}}}{M}} \right)}^{6n}} + \cdots} \right].\label{TH2}
\end{eqnarray}
It is obvious that the first term of the above result coincides with Hawking's result and the case $n=1$ covers the result given in the framework of the exponential GUP~\cite{s17}. Moreover, we provide the new temperature-mass relation for $n \ge 2$ in the framework of our improved exponential GUP~\cite{s18}. We note that when $n$ increases, the new temperature-mass relation is close to the Hawking result in the process before the end of evaporation, but quite different from the Hawking's at the end of evaporation in the aspects that the new relation leads to a finite maximal temperature 
and a non-vanishing minimal mass. 

Using the numerical method, we plot the curves of the Hawking temperature versus the black hole mass in Figures 1 and 2 in which the curves from the Hawking proposal, the quadratic GUP (noted by $\mathrm{GUP_0}$), and the exponential GUP (noted by $\mathrm{GUP_1}$) are shown together for comparisons. Note that we use the Planck units\footnote{Here we list some Planck units related to this work and their values in the SI units. $\ell_{\rm Pl} = 1.61620 \times10^{-35}\, m$, $M_{\rm Pl} =2.17651 \times10^{-8}\, kg$,  $T_{\rm Pl}=1.41683 \times 10^{32}\, K$, $t_{\rm Pl} = 5.39106  \times 10^{-44}\, s$, the Planck unit of entropy: $S_{\rm Pl} =1.38065 \times 10^{-23}\, J/K$, the Planck unit of heat capacity: $C_{\rm Pl} =1.38065 \times 10^{-23}\, J/K$, and the Planck unit of  power of radiation: $P_{\rm Pl}=3.62831\times10^{53}\, J/s$.} in all figures and tables of the present paper.

\begin{figure}[!htbp]
\centering
\includegraphics[height=8cm]{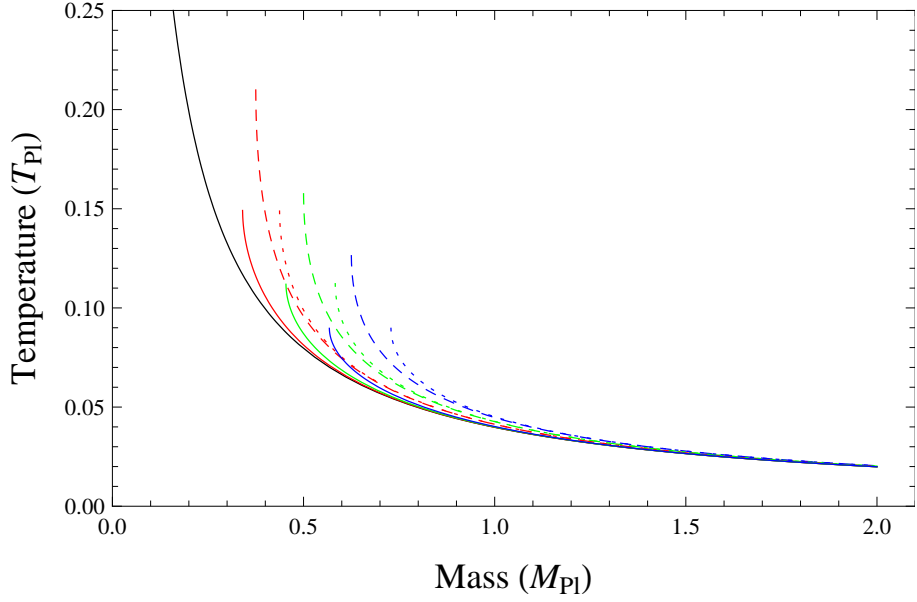}
\caption{The temperature versus the black hole mass for the case $n=2$. From left to right: the Hawking result (black solid curve), $\mathrm{GUP_2}$ result (solid curve), $\mathrm{GUP_0}$ result (dashed curve), and $\mathrm{GUP_1}$ result (dotted curve) for $\alpha=0.75$ (red), $\alpha=1$ (green), and $\alpha=1.25$ (blue), respectively.}
\end{figure}

Figure 1 shows that the maximal temperature decreases but the remnant mass  increases when the parameter $\alpha$ grows.

\begin{figure}[!htbp]
\centering
\includegraphics[height=8cm]{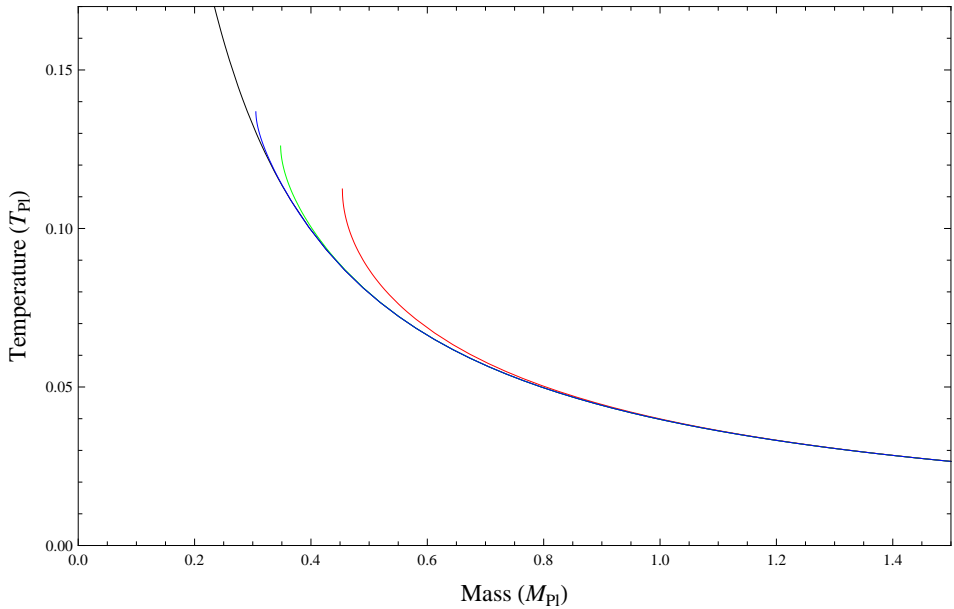}
\caption{The temperature versus the black hole mass for the case  $\alpha=1$. Curves are the Hawking result (black) and $\mathrm{GUP_n}$ results for $n=2$ (red), $n=5$ (green), and $n=10$ (blue), respectively.}
\end{figure}

Figure 2 shows that the maximal temperature increases but the remnant mass decreases when the index $n$ grows. Moreover,
the temperature-mass curve of $\mathrm{GUP_n}$ is close to the Hawking curve when $n$ grows, but the significant difference between the two cases is that the former is bounded by a finite maximal temperature and a non-vanishing minimal mass (BHR) at the final stage of evaporation due to the GUP quantum effects.

In addition, we can obtain the black hole mass as a function of the Hawking temperature through solving eq.~(\ref{TH}),
\begin{eqnarray}
M = \frac{1}{{8\pi TG}}\exp \left\{ {\frac{1}{{2n}}{{\left( {\frac{{{T}}}{{T_{\rm Max}}}} \right)}^{2n}}} \right\},\label{Mt}
\end{eqnarray}
where $T_{\rm Max} = \left( {\frac{1}{{2n}}} \right)^{1/(2n)}\frac{{{T_{\rm Pl}}}}{{2\pi \alpha }}$  is the maximal temperature the black hole can reach once the black hole mass reduces to the minimal value $M_0$. That is, $T_{\rm Max}$, approximately in the order of ten percent of the Planck temperature for any $n$, is the temperature of the black hole remnant. 
We note that it is a common property that $M_0$ is non-vanishing and  $T_{\rm Max}$ does not go to infinity in the framework of any GUP, which is different from the Hawking proposal. Incidentally, our result eq.~(\ref{Mt}) reduces to that of  ref.~\cite{s17} when $n=1$.

\subsection{Entropy}
Now we calculate the micro-canonical entropy of the Schwarzschild black hole. It is known~\cite{Scardigli} that the minimal increase of the area of a black hole when absorbing a classical particle is ${\left( {\Delta A} \right)_0} \approx 8\,\ell_{\rm Pl}^2\left( {\ln 2} \right)\left(\Delta X\right)\left(\Delta P\right)$. Using the saturate uncertainty relation (see eq.~(\ref{sur}) or eq.~(\ref{pmp})) and considering that the minimal horizon area and the horizon area can be expressed by $A_0 =4 \pi ({\Delta X})_0^2 $  and $A =4 \pi (\Delta X)^2 $, respectively, we give the minimal increase of the black hole area,
\begin{eqnarray}
{\left( {\Delta A} \right)_0} \approx 4\,\ell_{\rm Pl}^2\left( {\ln 2} \right)\exp \left\{ { - \frac{1}{{2n}}W\left( { - \frac{1}{e}{{\left( {\frac{{{A_0}}}{A}} \right)}^n}} \right)} \right\}.
\end{eqnarray}
Since the minimal increase of the entropy of a black hole is ${\left( {\Delta S} \right)_0} = \ln 2$, we approximately establish the following differential equation as usual,
\begin{eqnarray}
\frac{{dS}}{{dA}} \approx \frac{{{{\left( {\Delta S} \right)}_0}}}{{{{\left( {\Delta A} \right)}_0}}} = \frac{1}{{4\,\ell_{\rm Pl}^2}}\exp \left\{ { \frac{1}{{2n}}W\left( { - \frac{1}{e}{{\left( {\frac{{{A_0}}}{A}} \right)}^n}} \right)} \right\}.
\end{eqnarray}
As a result, considering the minimal horizon area $A_0$ as the lower limit of the horizon area integration, we give the entropy of the black hole as follows,
\begin{eqnarray}
S = \frac{1}{4\,\ell_{\rm Pl}^2}\int_{{A_0}}^A \exp \left\{ {\frac{1}{{2n}}W\left( { - \frac{1}{e}{{\left( {\frac{{{A_0}}}{A}} \right)}^n}} \right)} \right\}dA.
\end{eqnarray}
It is evident that our result reduces to that of the exponential GUP when $n=1$. For $n \ge 2$, we shall give a new entropy-area or entropy-mass relation by following a particular treatment in area integration.

Setting $y \equiv  - \frac{1}{e}{\left( {\frac{{{A_0}}}{A}} \right)^n}$ and using the property of the Lambert $W$ function: $\exp \left( {\frac{{W\left( u \right)}}{{2n}}} \right) = \left( {\frac{u}{{W\left( u \right)}}} \right)^{1/(2n)}$, we perform  the above integration and obtain the entropy as a function of the black hole area,
\begin{eqnarray}
S &= & \frac{{{A_0}}}{{4n\ell_{\rm Pl}^2{e^{\frac{1}{n}}}}}\int_{ - \frac{1}{e}}^{ - \frac{1}{e}{{\left( {\frac{{{A_0}}}{A}} \right)}^n}} {{\left( -y \right)^{ - 1 - \frac{1}{{2n}}}}} \left(-W{\left( y \right)}\right)^{ - \frac{1}{{2n}}}dy  \nonumber\\
&=& \frac{{{{\left( { - 1} \right)}^{ - \frac{1}{n}}}{A_0}}}{{4n\ell_{\rm Pl}^2{{\left( {2ne} \right)}^{\frac{1}{n}}}}}\Bigg\{2n\Gamma \left( {\frac{{n - 1}}{n},\frac{1}{{2n}}W\left( { - \frac{1}{e}{{\left( {\frac{{{A_0}}}{A}} \right)}^n}} \right)} \right) - 2n\Gamma \left( {\frac{{n - 1}}{n}, - \frac{1}{{2n}}} \right)\Bigg. \nonumber \\
& & + \Bigg.\Gamma \left( { - \frac{1}{n},\frac{1}{{2n}}W\left( { - \frac{1}{e}{{\left( {\frac{{{A_0}}}{A}} \right)}^n}} \right)} \right) - \Gamma \left( { - \frac{1}{n}, - \frac{1}{{2n}}} \right) \Bigg\}, \label{entropyarea}
\end{eqnarray}
or the entropy as a function of the black hole mass,
\begin{eqnarray}
S &= & \frac{{{{\left( { - 1} \right)}^{ - \frac{1}{n}}}{\pi \alpha^2}}}{{4n}}\Bigg\{2n\Gamma \left( {\frac{{n - 1}}{n},\frac{1}{{2n}}W\left( { - \frac{1}{e}{{\left( {\frac{{{M_0}}}{M}} \right)}^{2n}}} \right)} \right) - 2n\Gamma \left( {\frac{{n - 1}}{n}, - \frac{1}{{2n}}} \right)\Bigg. \nonumber \\
& & + \Bigg.\Gamma \left( { - \frac{1}{n},\frac{1}{{2n}}W\left( { - \frac{1}{e}{{\left( {\frac{{{M_0}}}{M}} \right)}^{2n}}} \right)} \right) - \Gamma \left( { - \frac{1}{n}, - \frac{1}{{2n}}} \right) \Bigg\}, \label{entropymass}
\end{eqnarray}
where the Cauchy principal value of the above integral has been chosen and the definition of the upper incomplete gamma function is $\Gamma(s,x) \equiv \int_x^{\infty}t^{s-1} e^{-t} dt$. Note that the entropy remains real even if the index $n$ is even.

As done in the above subsection, by using the numerical method, we plot the curves of the entropy versus the black hole mass in Figures 3 and 4 in which the curves from the Hawking proposal, the quadratic GUP (noted by $\mathrm{GUP_0}$), and the exponential GUP (noted by $\mathrm{GUP_1}$) are shown together for comparisons.

\begin{figure}[!htbp]
\centering
\includegraphics[height=8cm]{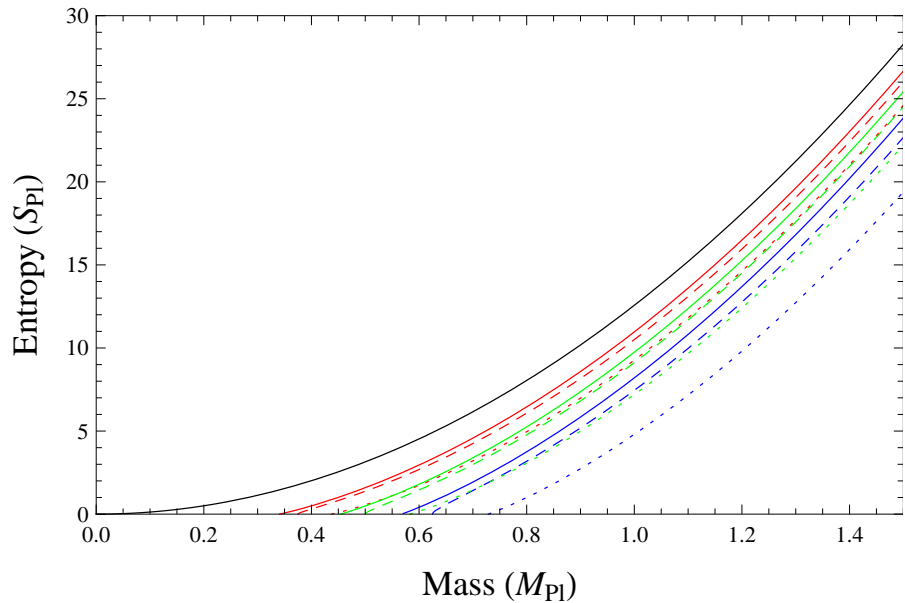}
\caption{The entropy versus the black hole mass for the case $n=2$. From left to right: the Hawking result (black solid curve), $\mathrm{GUP_2}$ result (solid curve), $\mathrm{GUP_0}$ result (dashed curve), and $\mathrm{GUP_1}$ result (dotted curve) for $\alpha=0.75$ (red), $\alpha=1$ (green), and $\alpha=1.25$ (blue), respectively.}
\end{figure}

Figure 3 indicates that when $\alpha$ is growing, the entropy of the black hole with a fixed mass is declining and the zero-entropy remnant has an increasing mass at the final stage of evaporation.

\begin{figure}[!htbp]
\centering
\includegraphics[height=8cm]{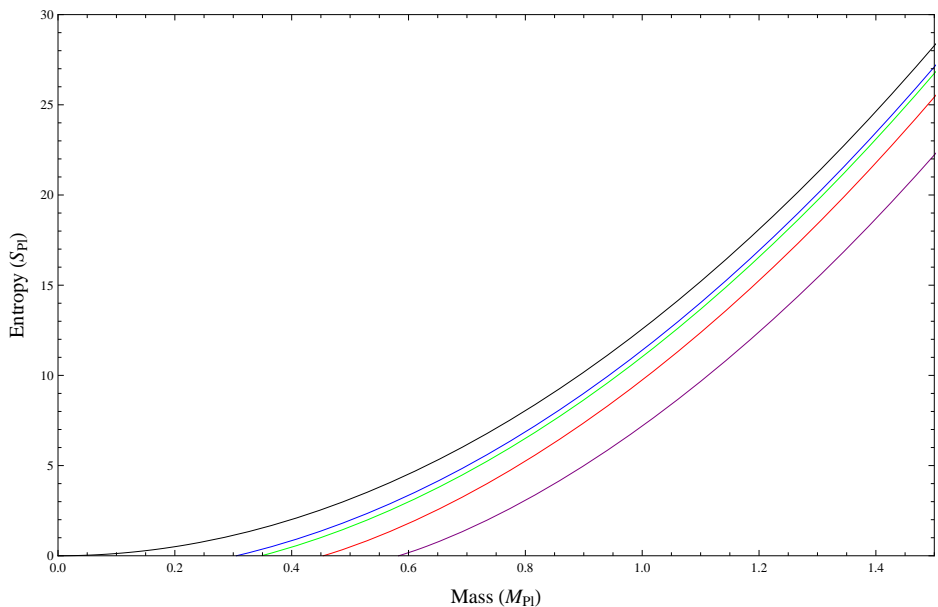}
\caption{The entropy versus the black hole mass for the case $\alpha=1$. Curves are the Hawking result (black) and $\mathrm{GUP_n}$ results for $n=2$ (red), $n=5$ (green), and $n=10$ (blue), respectively. Moreover, the $\mathrm{GUP_1}$ result (purple) is shown for comparison.}
\end{figure}

Figure 4 indicates that when $n$ is growing, the entropy of the black hole with a fixed mass is increasing and the zero-entropy remnant has a declining mass at the final stage of evaporation. Moreover,
the entropy-mass curve of $\mathrm{GUP_n}$ is close to the Hawking curve for a large $n$, but the former will not overlap the latter because the minimal mass (BHR) $M_0$ is non-vanishing. 

In addition, it may be of some interest to analyze how our entropy-area relation modifies the standard Bekenstein-Hawking entropy formula. To this end, we expand eq.~(\ref{entropyarea}) for the cases $n \ge 2$ in $\frac{1}{e}\left(A_0/A\right)^n$ up to the third order,
\begin{eqnarray}
S & \approx &\frac{A}{4\ell_{\rm Pl}^2}- \frac{{{A_0}}}{4\ell_{\rm Pl}^2}\left\{1 + \frac{1}{{2n\left( {n - 1} \right)e}}\left[1-{{\left( {\frac{{{A_0}}}{A}} \right)}^{n-1}}\right] + \frac{4n-1}{{8{n^2}\left( {2n - 1} \right){e^2}}}\left[1-{{\left( {\frac{{{A_0}}}{A}} \right)}^{2n-1}}\right]\right. \nonumber \\
& & \left.+\frac{{{{\left( {6n - 1} \right)}^2}}}{{48{n^3}\left( {3n - 1} \right){e^3}}}\left[1-{\left( {\frac{{{A_0}}}{A}} \right)^{3n-1}} \right]\right\}.\label{entropyareaapp}
\end{eqnarray}
We see that the leading order coincides with the Bekenstein-Hawking entropy formula, 
but the sub-leading order is a power-law correction $- \frac{{{A_0}}}{4\ell_{\rm Pl}^2}\left\{1 + \frac{1}{{2n\left( {n - 1} \right)e}}\left[1-{{\left( {\frac{{{A_0}}}{A}} \right)}^{n-1}}\right]\right\}$ instead of the well-known logarithmic correction, where $A_0$ is the minimal horizon area related to the black hole remnant that is vanishing in the framework of  the Hawking proposal but non-vanishing in the framework of  any GUP.  We also point out a common property that our correction and the logarithmic correction possess, that is, both are negative. 
Qualitatively, we can see from Figure 4 that the absolute value of our correction for any case of $n \ge 2$ is smaller than that of the logarithmic correction that corresponds to the case $n=1$. Further, the larger $n$ is, the smaller the absolute value of our correction goes to. Quantitatively, we can estimate the difference of the two distinct corrections. The absolute value of the correction in the  framework of $\mathrm{GUP_1}$ is
$|{\Delta {S_1}}(M)|  \approx   \frac{{\pi {\alpha ^2}}}{2}\ln \left( {\frac{M}{{{M_0}}}} \right) - \frac{{3\pi {\alpha ^2}}}{{16e}}{\left( {\frac{{{M_0}}}{M}} \right)^2} + \frac{{\pi {\alpha ^2}}}{4}\left[ { - \gamma  + 1 + 2\sqrt e  + {\rm Ei}\left( {\frac{1}{2}} \right) + \ln \left( {2e} \right)} \right]$, and in the framework of $\mathrm{GUP_n}$ it is $|{\Delta {S_n}}(M)| \approx  \frac{{\pi {\alpha ^2}}}{4}{\left( {2ne} \right)^{\frac{1}{n}}}\left\{ {1 + \frac{1}{{2n\left( {n - 1} \right)e}}\left[ {1 - {{\left( {\frac{{{M_0}}}{M}} \right)}^{2n - 2}}} \right]} \right\}$, where ${\rm Ei}(x)$ is the exponential integral function defined as ${\rm Ei}(x) \equiv -\int_{-x}^\infty  {t^{ - 1}} {e^{ - t}}dt$ and $\gamma$ is the Euler constant. We compute the ratio of the two corrections for three samples: (i) For $M = 2 {M_{\rm Pl}}$ and $n=2$, $\left|\frac{{\Delta {S_2}}(2)}{{\Delta {S_1}}(2)}\right| \approx 4.4 \times 10^{-1}$; (ii) For $M = 10 {M_{\rm Pl}}$ and $n=2$, $\left|\frac{{\Delta {S_2}}(10)}{{\Delta {S_1}}(10)}\right| \approx 3.2 \times 10^{-1}$; (iii) For $ M = 10 {M_{\rm Pl}}$ and  $n=100$, $\left|\frac{{\Delta {S_{100}}}(10)}{{\Delta {S_1}}(10)}\right| \approx 9.2 \times 10^{-2}$. It is evident that the quantitative results coincide with the qualitative analysis. That is to say, in general, the correction of $\mathrm{GUP_n}$ ($n \ge 2$) is smaller than that of $\mathrm{GUP_1}$ (logarithm) for the black holes with various masses and $n$'s;  to be specific, the greater the mass is, for a fixed $n$, the smaller the ratio is, and moreover, the greater $n$ is, for a fixed mass, the smaller the ratio becomes, which means that the deviation of the $\mathrm{GUP_n}$ correction from the logarithmic correction goes to greater.

\subsection{Heat capacity}
By using eq.~(\ref{Mt}), we get the heat capacity of the black hole,
\begin{eqnarray}
C &=& \frac{{dM}}{{d{T}}} \nonumber\\
& =&  - 8\pi {M^2}\ell_{\rm Pl}^{2}\left[ {1 + W\left( { - \frac{1}{e}{{\left( {\frac{{{M_0}}}{M}} \right)}^{2n}}} \right)} \right]\exp \left[ {\frac{1}{{2n}}W\left( { - \frac{1}{e}{{\left( {\frac{{{M_0}}}{M}} \right)}^{2n}}} \right)} \right] \nonumber\\
& = & - 8\pi {M^2}\ell_{\rm Pl}^{2}\left[ 1 - \left( {1 + \frac{1}{{2n}}} \right)\frac{1}{e}{{\left( {\frac{{{M_0}}}{M}} \right)}^{2n}} + \left( {\frac{1}{{8{n^2}}} - 1} \right)\frac{1}{{{e^2}}}{{\left( {\frac{{{M_0}}}{M}} \right)}^{4n}} \right.\nonumber\\
& & \left.+ \frac{{ - 1 + 6n + 12{n^2} - 72{n^3}}}{{48{n^3}}}\frac{1}{{{e^3}}}{{\left( {\frac{{{M_0}}}{M}} \right)}^{6n}} + \cdots \right],\label{capacity}
\end{eqnarray}
which shows that the heat capacity vanishes at the end point of evaporation when $M=M_0$, i.e. when $M$ equals the mass of black hole remnant. We describe the variation of the heat capacity with the black hole mass in Figures $5$ and $6$.

\begin{figure}[!htbp]
\centering
\includegraphics[height=8cm]{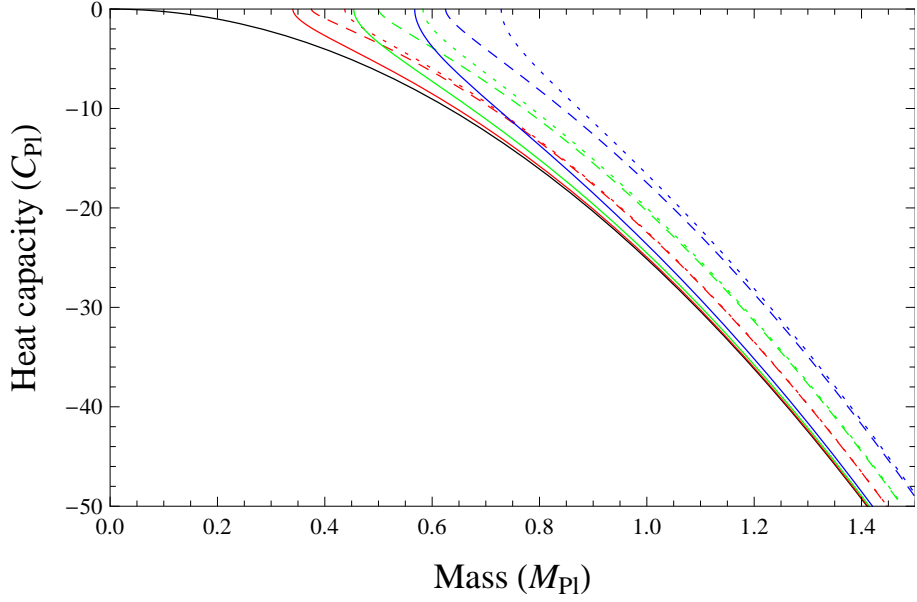}
\caption{ The heat capacity versus the black hole mass for the case $n=2$. From left to right: the Hawking result (black solid curve), $\mathrm{GUP_2}$ result (solid curve), $\mathrm{GUP_0}$ result (dashed curve), and $\mathrm{GUP_1}$ result (dotted curve) for $\alpha=0.75$ (red), $\alpha=1$ (green), and $\alpha=1.25$ (blue), respectively.}
\end{figure}

Figure 5 indicates that when $\alpha$ is growing, the heat capacity of the black hole with a fixed mass is increasing (but its absolute value is declining) and the remnant with vanishing heat capacity also has an increasing mass at the end point of evaporation.

\begin{figure}[!htbp]
\centering
\includegraphics[height=8cm]{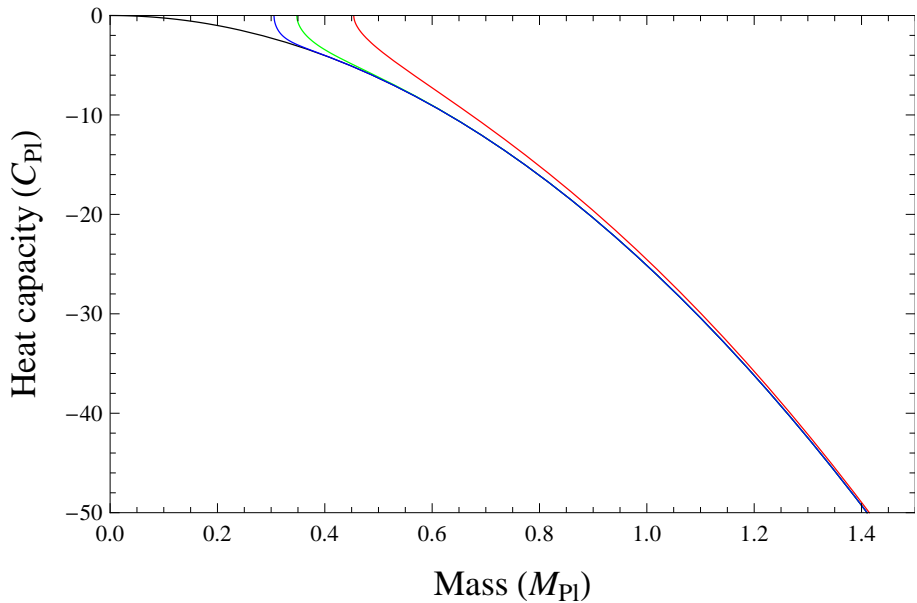}
\caption{ The heat capacity versus the black hole mass for the case  $\alpha=1$. Curves are the Hawking result (black) and $\mathrm{GUP_n}$ results for $n=2$ (red), $n=5$ (green), and $n=10$ (blue), respectively.}
\end{figure}

Figure 6 indicates that when $n$ is growing, the heat capacity of the black hole with a fixed mass is declining (but its absolute value is increasing) and the remnant with vanishing heat capacity also has a declining mass at the end point of evaporation.


\section{Black hole evaporation}
In this section we focus on the evaporation of the Schwarzchild black hole in a variety of GUP frameworks. Supposing that black holes
only emit photons, we give the evaporation rate and the decay time without and with the consideration of the UV/IR mixing effect, respectively. We shall see that the UV/IR mixing effect plays a significant role in the novel understanding of the Hawking radiation from the point of view of GUPs. That is, this effect can largely slow down the rate of evaporation. 
As the results cannot be expressed analytically, we thus list them numerically in the following investigations.

\subsection{Black hole evaporation without the consideration of the UV/IR mixing effect}
In this subsection we follow the usual way, see, for instance, ref.~\cite{s17} where the UV/IR mixing effect is not considered, to investigate the black hole evaporation.
We know that the weighted phase space volume ${{e^{ - 3 {\alpha ^{2n}}\ell_{\rm Pl}^{2n}{P^{2n}}}}} {d^3}{\bf X} {d^3}{\bf P}$, where
$P=\sqrt{{\bf P} \cdot {\bf P}}$, is invariant under time evolution, which is known as the analog of the Liouville theorem in the classical limit~\cite{s18}. Therefore, the density of states in the momentum space has the form $ \frac{1}{{{{\left( {2 \pi } \right)}^3}}} {{e^{ - 3 {\alpha ^{2n}}\ell_{\rm Pl}^{2n}{P^{2n}}}}} {d^3}{\bf P} $ and the average energy per volume at temperature $T$ reads as
\begin{eqnarray}
\mathcal{E}_\gamma (T)=  \frac{2}{{{{\left( {2\pi } \right)}^3}}} \int {{e^{ - 3 {\alpha ^{2n}}\ell_{\rm Pl}^{2n}{P^{2n}}}}} \frac{P{d^3}{\bf P} }{{{e^{P/T}}} - 1}.\label{aepv}
\end{eqnarray}
According to the Stefan-Boltzmann law, we give the evaporation rate\footnote{Consider photons radiating out from a black hole via an infinitesimal surface $dA$ whose solid angle is $d\Omega$. The evaporation energy per unit time is equal to  $\frac{{{{\cal E}_\gamma }(T)}}{{4\pi }}\cos \theta d\Omega dA$.
So the total evaporation rate can be calculated by integrating the solid angle over the half-sphere and a factor 1/4 emerges, that is, $\frac{{dM}}{{dAdt}} =  - \frac{{{{\cal E}_\gamma }(T)}}{{4\pi }}\int {\cos \theta } d\Omega  =  - \frac{{{{\cal E}_\gamma }(T)}}{{4\pi }}\int_0^{\pi /2} {\cos \theta \sin \theta d\theta } \int_0^{2\pi } {d\varphi }  =  - \frac{1}{4}{{\cal E}_\gamma }(T)$. The factor 1/4 was lost in ref.~\cite{s17}.} as follows,
\begin{eqnarray}
\frac{{dM}}{{dt}} = - \frac{A}{4}{\mathcal{E}_\gamma }(T),\label{er}
\end{eqnarray}
where $A$ is the horizon area. Substituting eq.~(\ref{aepv}) into eq.~(\ref{er}) and making the momentum integration from zero as the lower limit, we obtain
\begin{eqnarray}
\frac{{dM}}{{dt}} =   - \frac{{4{G^2}{M^2}}}{\pi } \int_{0}^\infty  {{e^{-3 {\alpha ^{2n}}\ell_{\rm Pl}^{2n}{P^{2n}}}}\frac{{{P^3}dP}}{{{e^{{P/T}}} - 1}}}. \label{dMt}
\end{eqnarray}
Further considering the temperature as the function of the mass, eq.~(\ref{TH}), we then draw the pictures to show the relations between the evaporation rate and the black hole mass in Figures $7$ and $8$. Note that the UV/IR mixing effect is not embedded in eq.~(\ref{dMt}) because that zero is taken as the lower limit implies that the sub-Planckian modes are not excluded~\cite{Bank, s19,s18} in the contribution to the energy density.

\begin{figure}[!htbp]
\centering
\includegraphics[height=8cm]{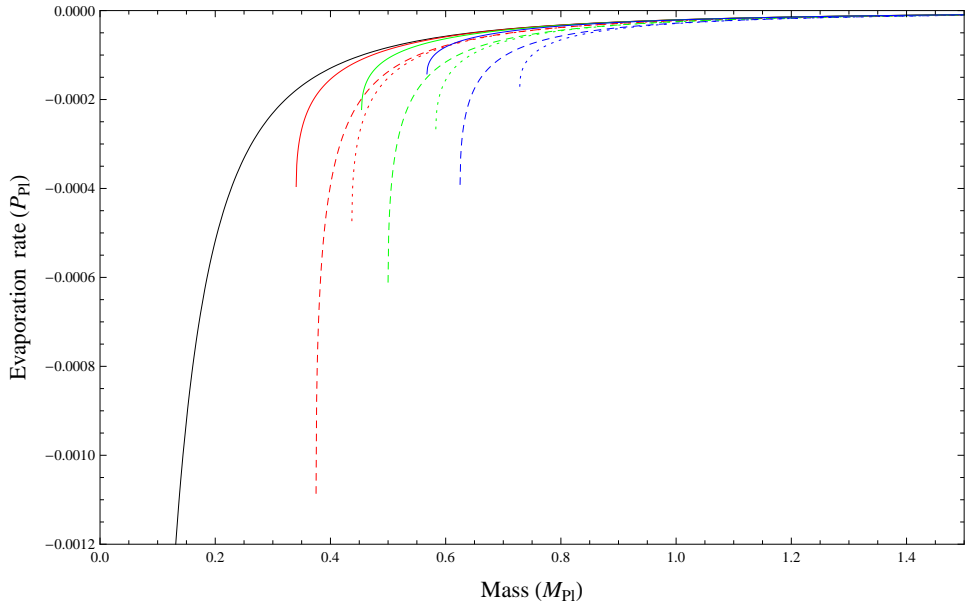}
\caption{The evaporation rate versus the black hole mass for the case $n=2$. From left to right: the  Hawking result (black solid curve), $\mathrm{GUP_2}$ result (solid curve), $\mathrm{GUP_0}$ result (dashed curve), and $\mathrm{GUP_1}$ result (dotted curve)   for $\alpha=0.75$ (red), $\alpha=1$ (green), and $\alpha=1.25$ (blue), respectively.}
\end{figure}

\begin{figure}[!htbp]
\centering
\includegraphics[height=8cm]{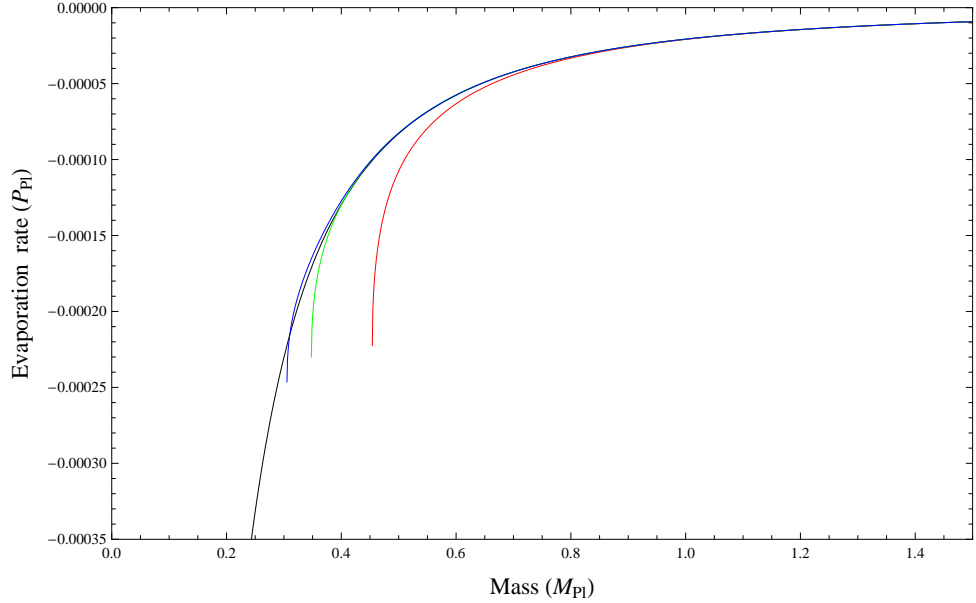}
\caption{The evaporation rate versus the black hole mass for the case $\alpha=1$. Curves are the  Hawking result (black), $\mathrm{GUP_n}$ results for $n=2$ (red), $n=5$ (green), and $n=10$ (blue), respectively.}
\end{figure}

Figure 7 shows that for a black hole with a fixed mass the absolute value of the evaporation rate in the framework of any of the three GUPs is greater than that of the Hawking proposal, which means that the radiation process speeds up in the former case. Further, the larger $\alpha$ is, the stronger the speedup of evaporation becomes. Moreover, the parameter $\alpha$ and the remnant mass have a positive correlation.

Figure 8 shows that for a black hole with a fixed mass the absolute value of the evaporation rate in the framework of $\mathrm{GUP_n}$  with small $n$, like $n=2, 5$, is greater than that of the Hawking proposal, which means that the radiation process speeds up in the former case. When $n$ becomes larger, like $n=10$, the curve of $\mathrm{GUP_{10}}$ has two points of intersection with that of the Hawking proposal, which implies that the radiation slows down in the framework of $\mathrm{GUP_{10}}$ when the black hole mass takes the values of interval corresponding to the two points. Moreover, the index $n$ and the remnant mass have a negative correlation.


The decay time of the evaporation process can be calculated by integrating eq.~(\ref{dMt}) with respect to the black hole mass  from $M$ to $M_0$,
\begin{eqnarray}
t = -\frac{\pi}{4{G^2}}\int_{M}^{M_0} {\frac{{dM}}{{ {M^2}\int_{0}^\infty  {{e^{-3 {\alpha ^{2n}}\ell_{\rm Pl}^{2n}{P^{2n}}}}\frac{{{P^3}dP}}{{{e^{{P/T}}} - 1}}} }}}.
\end{eqnarray}
Note that the evaporation process ends when the black hole mass reaches the minimal mass $M_0$. Using the relation between the temperature and the mass, see eq.~(\ref{TH}),
we compute numerically the decay time as the function of the black hole mass and give the Hawking decay time, the $\mathrm{GUP_0}$-, $\mathrm{GUP_1}$-, and $\mathrm{GUP_n}$-corrected decay times in the following two tables.

\begin{table}[!htbp]
\small
\centering
\begin{tabular}{|c|*{6}{|c}}
\hline
\multicolumn{6}{|c|} {Hawking time, and $\mathrm{GUP_0}$- and $\mathrm{GUP_1}$-corrected decay times without the UV/IR mixing effect} \\ \hline
\backslashbox{Frame}{M}
& $1$ & $2$ & $3$ & $4$ & $5$  \\ \hline \hline
Hawking & $16085$ & $128680$ & $434294$ & $1.02944\times10^6$ & $2.01062\times10^6 $\\ \hline
$\mathrm{GUP_0}$ & $9838.03 $&$ 114511 $&$412314 $&$ 999674 $&$ 1.97308\times10^6$ \\ \hline
$\mathrm{GUP_1}$ &$ 9070.31 $&$ 113601 $&$ 411361 $&$ 998701 $&$ 1.97210\times10^6$  \\ \hline
\end{tabular}
\caption{Hawking time, $\mathrm{GUP_0}$- and $\mathrm{GUP_1}$-corrected decay times with the black hole mass $M=1, 2, \cdots, 5$ (in Planck units) and $\alpha=1$.}
\end{table}

\begin{table}[!htbp]
\small
\centering
\begin{tabular}{|c|*{6}{|c}}
\hline
\multicolumn{6}{|c|} {$\mathrm{GUP_n}$-corrected decay time without the UV/IR mixing effect} \\ \hline
\backslashbox{n}{M}
& $1$ & $2$ & $3$ & $4$ & $5$  \\ \hline \hline
2 & $13868.8$&$126225$&$ 431761$&$ 1.02687\times10^6$&$2.00802\times10^6$\\ \hline
3 & $14911.3$&$127502$&$ 433115$&$ 1.02826\times10^6$&$ 2.00944\times10^6 $  \\ \hline
4 & $15248.1$&$ 127843$&$ 433457$&$ 1.0286\times10^6$&$ 2.00978\times10^6$  \\ \hline
5 & $15405$&$ 128000$&$ 433614$&$ 1.02876\times10^6$&$ 2.00994\times10^6$  \\ \hline
6 & $15495.8$&$ 128091$&$ 433705$&$ 1.02885\times10^6$&$ 2.01003\times10^6$ \\ \hline
7 & $15555.3$&$ 128150$&$ 433764$&$1.02891\times10^6$&$ 2.01009\times10^6$ \\ \hline
8 & $15597.5$&$ 128192$&$ 433806$&$1.02895\times10^6$&$ 2.01013\times10^6$ \\ \hline
9 & $ 15629.1$&$ 128224$&$433838$&$ 1.02898\times10^6$&$ 2.01016\times10^6$ \\ \hline
10 & $15653.7$&$ 128248$&$ 433863$&$ 1.02901\times10^6$&$ 2.01019\times10^6$ \\ \hline
\end{tabular}
\caption{ $\mathrm{GUP_n}$-corrected decay time with $n=2, 3, \cdots,10$, the black hole mass $M=1, 2, \cdots, 5$ (in Planck units) and $\alpha=1$.}
\end{table}

Table 1 indicates that the decay time\footnote{The numerical values in ref.~\cite{s17} are incorrect, we correct them in the above table. The reason is that the coefficients $1/(2\pi)^3$ and $1/4$ were lost in eqs.~(36) and (44) of ref.~\cite{s17}, respectively.  See our eqs.~(\ref{aepv}) and (\ref{er}).} is longer for a larger (heavier) black hole in any framework of the Hawking, $\mathrm{GUP_0}$, and $\mathrm{GUP_1}$, which is usually reasonable. Due to the speedup of evaporation in the framework of $\mathrm{GUP_0}$ or $\mathrm{GUP_1}$, the decay time is shorter than that of the Hawking proposal for a fixed mass, but the deviation is small. Moreover, because the speedup of evaporation in the framework of $\mathrm{GUP_0}$ is weaker than that in the framework of $\mathrm{GUP_1}$, the decay time in the former is longer than that in the latter and the difference of decay times between the two GUPs is small for a fixed mass.

Table 2 indicates that the decay time is longer for a larger (heavier) black hole in the framework of $\mathrm{GUP_n}$ for any of $n \ge 2$ cases, which is same as the situation appeared in the Hawking proposal, $\mathrm{GUP_0}$, and $\mathrm{GUP_1}$. For a fixed mass, for instance, $M=2M_{\rm Pl}$, the decay time and the index $n$ have a positive correlation, but the difference of decay times between the two cases with distinct indices is small. 

\subsection{Black hole evaporation with the consideration of the UV/IR mixing effect}

In general, the UV/IR mixing means that a large momentum measurement precision $\Delta P$ (UV) corresponds to a large position measurement precision $\Delta X$ (IR). (Heisenberg uncertainty principle, (${\Delta X}) ({\Delta P}) \geq \hbar/2$, 
shows that a large $\Delta P$ (UV) corresponds to a small $\Delta X$ (UV).) Specifically, we discover~\cite{s18} that a GUP provides effectively an IR cutoff due to the UV/IR mixing. (Note that a natural UV cutoff is provided by the deformation factor of a GUP.) The Heisenberg uncertainty relation gives the fact that the position and momentum spaces are Fourier transforms of each other. The more to localize a wave packet in position space (smaller $\Delta X$) corresponds to the more to superimpose momentum states (larger ${\Delta P}$). In the usual case, there is no lower bound to $\Delta X$. As asserted in refs.~\cite{Bank,s19}, one can compress the wave packet as small as possible by simply superimposing states with ever larger momentum (ever shorter wavelength) to cancel out the tails of the position space distributions. The uncertainty relation of GUPs, see, for instance, eq.~(\ref{uncertainty1}) or eq.~(\ref{uncertainty2}), implies that when one keeps on superimposing states with momenta larger than $(\Delta P)_{\rm Crit}$, $\Delta X$ stops decreasing and starts increasing instead. The natural interpretation of such a phenomenon would be that when the trans-Planckian modes (the states with momenta larger than $(\Delta P)_{\rm Crit}$) are superimposed on the sub-Planckian modes (the states with momenta smaller than $(\Delta P)_{\rm Crit}$), the trans-Planckian modes would ``jam" the sub-Planckian modes and prevent them from canceling out the tails of the wave packets effectively, i.e. the sub-Planckian modes are suppressed by the trans-Planckian modes. This brings about a shift of lower limit of momentum integral from zero to the critical momentum $(\Delta P)_{\rm Crit}$ which is dealt with as an effective IR cutoff.

In this subsection the UV/IR mixing effect is considered in the calculations of the evaporation rate and the decay time. The way to introduce this effect, as briefly explained above, is to exclude the sub-Planckian modes in the contribution to the energy density. Consequently, the lower limit of the momentum integration in eq.~(\ref{aepv}) should not be zero but the critical momentum (eq.~(\ref{Cri})) that corresponds to the minimal length. So, the evaporation rate of the black hole takes the form,
\begin{eqnarray}
\frac{{dM}}{{dt}} =   - \frac{{4{G^2}{M^2}}}{\pi} \int_{{\left( {\frac{1}{{2n}}} \right)^{{1}/{{(2n)}}}}\frac{1}{{\alpha \ell_{\rm Pl}}}}^\infty  {{e^{-3 {\alpha ^{2n}}\ell_{\rm Pl}^{2n}{P^{2n}}}}\frac{{{P^3}dP}}{{{e^{{P/T}}} - 1}}}. \label{dMt2}
\end{eqnarray}
Again using eq.~(\ref{TH}) that describes the Hawking temperature as a function of the black hole mass, we can plot the relations between the evaporation rate and the black hole mass in Figures 9 and 10, where the UV/IR mixing effect gives rise to a great deviation from the Hawking curve.

\begin{figure}[!htbp]
\centering
\includegraphics[height=8cm]{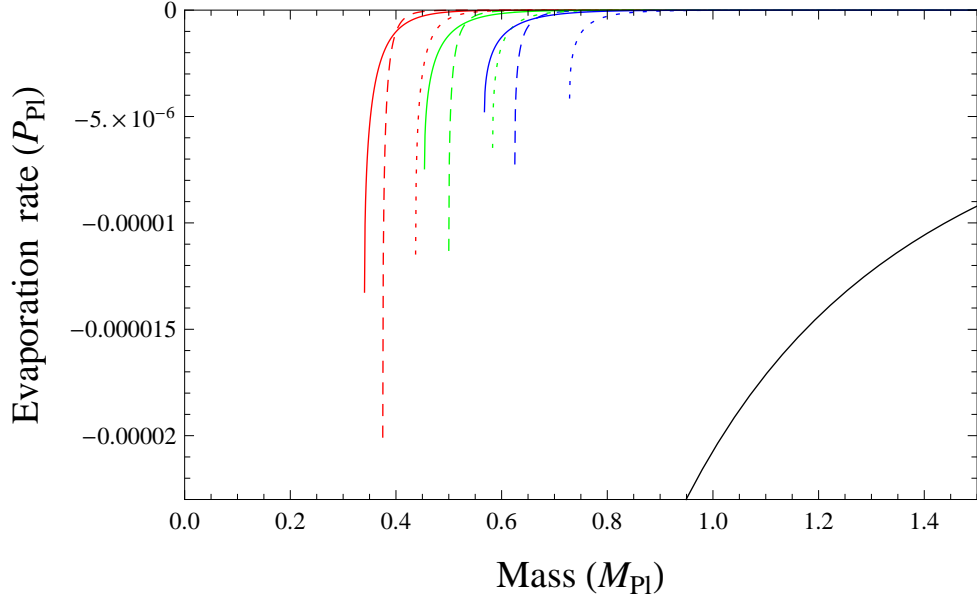}
\caption{The evaporation rate versus the black hole mass for the case $n=2$. From left to right: $\mathrm{GUP_2}$ result (solid curve), $\mathrm{GUP_0}$ result (dashed curve), and $\mathrm{GUP_1}$ result (dotted curve) for $\alpha=0.75$ (red), $\alpha=1$ (green), and $\alpha=1.25$ (blue), respectively, and the Hawking result (black solid curve) at the lower right corner.}
\end{figure}

\begin{figure}[!htbp]
\centering
\includegraphics[height=8cm]{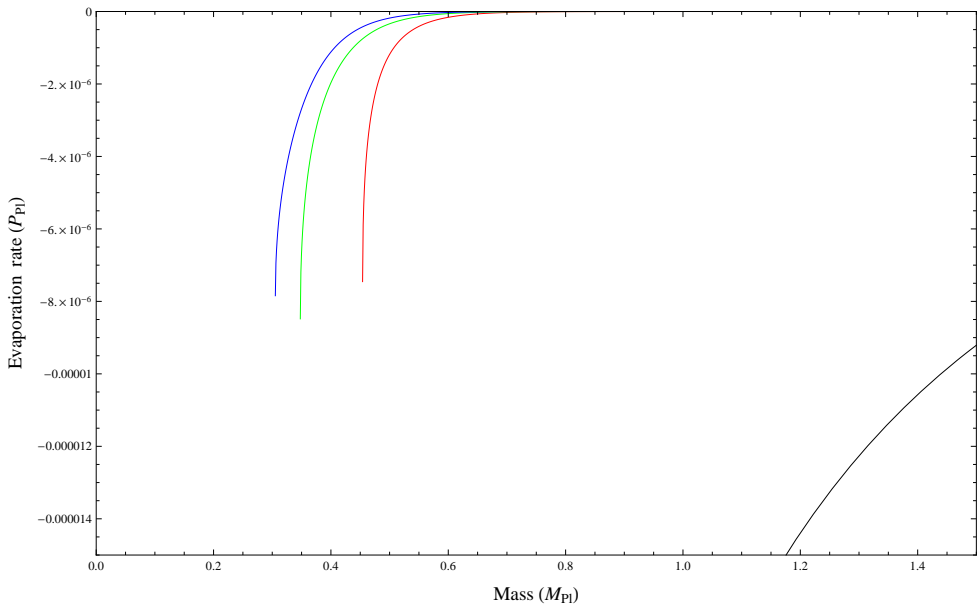}
\caption{ The evaporation rate versus the black hole mass for the case $\alpha=1$. Curves are the $\mathrm{GUP_n}$ results for $n=2$ (red), $n=5$ (green), and $n=10$ (blue), respectively, and the Hawking result (black) at the lower right corner.}
\end{figure}

Figure 9 means that for a black hole with a fixed mass the absolute value of the evaporation rate in the framework of any of the three GUPs is greatly slowed down by the UV/IR mixing effect, and therefore it is much smaller than that of the Hawking proposal, which gives a quite different situation from that where the UV/IR mixing effect is omitted. Moreover, the parameter $\alpha$ and the absolute value of the evaporation rate have a positive correlation, so do $\alpha$ and the remnant mass.

Figure 10 means that for a black hole with a fixed mass the absolute value of the evaporation rate in the framework of $\mathrm{GUP_n}$ with $n \ge 2$ is greatly slowed down by the UV/IR mixing effect, and therefore it is much smaller than that of the Hawking proposal, which gives a quite different situation from that where the UV/IR mixing effect is omitted. Moreover, the index $n$ and the absolute value of the evaporation rate have a negative correlation, so do $n$ and the remnant mass.

From eq.~(\ref{dMt2}), together with eq.~(\ref{TH}), we can easily obtain the decay time by integrating the black hole mass from $M$ to $M_0$,
\begin{eqnarray}
t = -\frac{\pi}{4{G^2}}\int_{M}^{M_0} {\frac{{dM}}{{ {M^2}\int_{{\left( {\frac{1}{{2n}}} \right)^{{1}/{{(2n)}}}}\frac{1}{{\alpha \ell_{\rm Pl}}}}^\infty  {{e^{-3 {\alpha ^{2n}}\ell_{\rm Pl}^{2n}{P^{2n}}}}\frac{{{P^3}dP}}{{{e^{{P/T}}} - 1}}} }}},
\end{eqnarray}
and give the results numerically in Tables 3 and 4. We can see that the decay time is largely prolonged in all cases relevant to GUP, which is definitely caused by the UV/IR mixing effect.

\begin{table}[!htbp]
\small
\centering
\begin{tabular}{|c|*{6}{|c}}
\hline
\multicolumn{6}{|c|} {Hawking time, and $\mathrm{GUP_0}$- and $\mathrm{GUP_1}$-corrected decay times with the UV/IR mixing effect} \\ \hline
\backslashbox{Frame}{M}
& $1$ & $2$ & $3$ & $4$ & $5$  \\ \hline \hline
Hawking & $16085$ & $128680$ & $434294$ & $1.02944\times10^6$ & $2.01062\times10^6 $ \\ \hline
$\mathrm{GUP_0}$ & $8.55644\times10^{10}$ & $ 9.44548\times10^{21}$ & $ 6.86861\times10^{32}$ & $ 4.85234\times10^{43}$ & $ 3.45809\times10^{54}$ \\ \hline
$\mathrm{GUP_1}$ & $ 1.9191\times10^8$ & $ 1.08681\times10^{16}$ & $ 4.63509\times10^{23}$ & $ 1.99868\times10^{31}$ & $ 8.83542\times10^{38}$  \\ \hline
\end{tabular}
\caption{Hawking time, $\mathrm{GUP_0}$- and $\mathrm{GUP_1}$-corrected decay times with the black hole mass $M=1, 2, \cdots, 5$ (in Planck units) and $\alpha=1$.}
\end{table}

\begin{table}[!htbp]
\small
\centering
\begin{tabular}{|c|*{6}{|c}}
\hline
\multicolumn{6}{|c|} {$\mathrm{GUP_n}$--corrected decay time with the UV/IR mixing effect} \\ \hline
\backslashbox{n}{M}
& $1$ & $2$ & $3$ & $4$ & $5$  \\ \hline \hline
2 & $3.51206\times 10^8 $&$ 9.35241\times 10^{15} $&$ 3.22915\times 10^{23} $&$ 1.25875\times 10^{31} $&$ 5.24436\times 10^{38}$ \\ \hline
3 &  $5.97713\times 10^8 $&$ 3.48826\times 10^{16} $&$ 2.85769\times 10^{24} $&$ 2.66066\times 10^{32} $&$ 2.6514\times 10^{40}$ \\ \hline
4 & $9.62743\times 10^8 $&$ 1.15183\times 10^{17} $&$ 1.96492\times 10^{25} $&$ 3.81531\times 10^{33} $&$ 7.93206\times 10^{41}$ \\ \hline
5 &  $1.44373\times 10^9 $&$ 3.05941\times 10^{17} $&$ 9.33582\times 10^{25} $& $3.2474\times 10^{34} $& $1.20992\times 10^{43}$ \\ \hline
6 &  $2.02961\times 10^9 $&$ 6.79022\times 10^{17} $&$ 3.3019\times 10^{26} $& $1.83312\times 10^{35} $& $1.09052\times 10^{44}$ \\ \hline
7 &  $2.70789\times 10^9 $&$ 1.31352\times 10^{18} $&$ 9.34557\times 10^{26} $& $7.60396\times 10^{35} $&$ 6.63259\times 10^{44}$ \\ \hline
8 &  $3.46654\times 10^9 $&$ 2.2875\times 10^{18} $&$ 2.2338\times 10^{27} $& $2.49888\times 10^{36} $& $2.99818\times 10^{45}$ \\ \hline
9 &  $4.29459\times 10^9 $&$ 3.67207\times 10^{18} $&$ 4.68656\times 10^{27} $&$ 6.86424\times 10^{36} $&$ 1.07885\times 10^{46}$ \\ \hline
10 & $5.18231\times 10^9 $&$ 5.52793\times 10^{18} $&$ 8.87472\times 10^{27} $& $1.63812\times 10^{37} $&$ 3.24633\times 10^{46}$ \\ \hline
\end{tabular}
\caption{ $\mathrm{GUP_n}$-corrected decay time with $n=2, 3, \cdots,10$, the black hole mass $M=1, 2, \cdots, 5$ (in Planck units) and $\alpha=1$.}
\end{table}

Table 3 indicates that  the decay time is longer for a larger (heavier) black hole in any framework of the Hawking, $\mathrm{GUP_0}$, and $\mathrm{GUP_1}$, which is usually reasonable. Due to a great slowdown of evaporation caused by the UV/IR mixing effect in the framework of $\mathrm{GUP_0}$ or $\mathrm{GUP_1}$, the decay time is much longer than that of the Hawking proposal for a fixed mass, and the deviation is huge. Moreover, because the slowdown of evaporation in the framework of $\mathrm{GUP_0}$ is stronger than that in the framework of $\mathrm{GUP_1}$, the decay time in the former is longer than that in the latter and the difference of decay times between the two GUPs is big for a fixed mass.

Table 4 indicates that the decay time is longer for a larger (heavier) black hole in the framework of $\mathrm{GUP_n}$ for any of $n \ge 2$ cases, which is same as the situation appeared in the Hawking proposal, $\mathrm{GUP_0}$, and $\mathrm{GUP_1}$. For a fixed mass, for instance, $M=2M_{\rm Pl}$, the decay time and the index $n$ have a positive correlation, and the difference of decay times between the two cases with distinct indices is big. 


\section{Conclusion}

In this paper we derive the maximally localized states for our improved exponential GUP, denoted by $\mathrm{GUP_n}$ with $n \ge 2$, and analyze some interesting properties of the states, such as the non-orthogonality, the corresponding quasi-position wavefunctions, and the scalar product of these wavefunctions. In addition, we investigate thermodynamics of the Schwardzschild black hole, i.e. we  actually calculate in the $\mathrm{GUP_n}$ framework the quantum corrections to some important thermodynamic quantities associated with the black hole, such as the Hawking temperature, the entropy, the heat capacity, the evaporation rate, the decay time, and the remnant mass. These results are summarized in the ten figures and four tables above.

We note that one can calculate the black hole thermodynamics for any function $f(\hat{P})$ when eq.~(\ref{XandP}) is generalized to be $[ {\hat{X}, \hat{P}} ] = i\hbar f(\hat{P})$. However, our improved exponential GUP is chosen non-trivially, which is demonstrated as follows. First of all, by using this GUP together with the consideration of the UV/IR mixing effect, we have given an interpretation of the Cosmological Constant Problem, see our previous work~\cite{s18} for the details. 
Further, we study in the present paper the black hole thermodynamics under the framework of our GUP, and indeed obtain  some new and interesting results. For example, our GUP modifies the maximally localized states and the Hawking evaporation of black holes. In particular, when the UV/IR mixing effect is involved, this GUP may radically change the fate of the black hole under evaporation. Although these results are all theoretical, they may provide us some new insights into old problems.

Finally, we summarize that the novelty of the present paper lies on two aspects for the black hole thermodynamics. One is that the entropy contains a power-law instead of logarithmic correction in the $\mathrm{GUP_n}$ framework, see eq.~(\ref{entropyareaapp}). The other aspect is that the evaporation rate and the decay time are computed with the consideration of the UV/IR mixing effect and that the two quantities greatly deviate from that obtained without introducing such an effect, see Tables 1-4 for the details, which implies that the UV/IR mixing effect produces a radical influence rather than a tiny correction to the black hole radiation. For instance, in the case $M=5M_{\rm Pl}$ and $n=10$ we compare the decay times in Table 2 without the UV/IR mixing effect and in Table 4 with such an effect, and see that the former is 40 orders of magnitude smaller than the latter. This means that the UV/IR mixing effect largely prolongs the radiation process of black holes. In particular,  in this case ($M=5M_{\rm Pl}$ and $n=10$) the decay time in Table 4 with the UV/IR mixing effect is in the order of $10^2$ seconds, which gives a quite available value if it is possible to be measured in future.
In addition, the difference between our GUP and the quadratic or exponential GUP exists in the higher (than second) order terms in the Taylor expansion. From Figures 7 and 9, one can see on the aspect of evaporation rates that the difference becomes apparent in the regime of small masses. Alternatively, the difference tends to be more apparent from the point of view of decay times if comparing the data of Table 1 with that of Table 2, or the data of Table 3 with that of Table 4. However, the Hawking radiation as the basis of our model has not yet been tested in experiment. The reason is obvious because the required high energy scale cannot be reached in laboratory and no specific signals of the Hawking radiation are observed in astronomy. Therefore, at present we have to leave the test of our results to far future experiments in laboratory, or in astronomical observations because we may infer that the evidences for testing thermodynamics of black holes might be hidden in primordial relics of the Big Bang.




\section*{Acknowledgments}
Y-GM would like to thank
H.P. Nilles of the University of Bonn for kind hospitality.
This work was supported in part by the Alexander von Humboldt Foundation under a renewed research program, by the National Natural
Science Foundation of China under grant No.11175090 and
by the Ministry of Education of China under grant No.20120031110027.

\end{document}